\documentclass[usenatbib]{mn2e}
\usepackage{longtable}
\input psfig.sty

\title[Binary Population Synthesis at High Spectral Resolution]
{Evolutionary Population Synthesis for Binary Stellar Population
at High Spectral Resolution: Integrated Spectral Energy
Distributions and Absorption-feature Indices}

\author[F. Zhang, L. Li and Z. Han]
{Fenghui~Zhang,\thanks{E-mail: gssephd@public.km.yn.cn;
zhang\_fh@hotmail.com} Lifang~Li and Zhanwen~Han\\
National Astronomical Observatories/Yunnan Observatory, Chinese
Academy of Sciences, PO Box 110, Kunming, \\ Yunnan Province,
650011, China \\}

\begin{document}

\date{\today}

\pagerange{\pageref{firstpage}--\pageref{lastpage}}

\pubyear{2005}

\maketitle

\label{firstpage}

\begin{abstract}
Using evolutionary population synthesis we present high resolution
($0.3\, \rm \AA$) integrated spectral energy distributions from
3000 to $7000\, \rm \AA$ and absorption-line indices defined by
the Lick Observatory image dissector scanner (referred to as the
Lick/IDS) system, for an extensive set of instantaneous burst
binary stellar populations with binary interactions. The ages of
the populations are in the range $1 - 15$\,Gyr and the
metallicities are in the range $0.004 - 0.03$. This high
resolution synthesis results can satisfy the needs of modern
spectroscopic galaxy surveys, and are available on request.

By comparing the synthetic continuum of populations at high and
low resolution we show that there is a good agreement for solar
metallicity and a tolerable disagreement for non-solar
metallicity. The strength of the Balmer lines at high spectral
resolution is greater than that at low resolution for all
metallicities. The comparison of Lick/IDS absorption-line indices
at low and high resolution, both of which are obtained by the
fitting functions, shows that the discrepancies in all indices
except for $\rm TiO_1$ and $\rm TiO_2$ are insignificant for
populations with $Z=0.004$ and $Z=0.02$. The high resolution
Ca4227, Fe5015 and $\rm Mg_b$ indices are redder than the
corresponding low resolution one for populations with $Z=0.01$ and
$Z=0.03$, this effect lowers the derived age and metallicity of
the population. The high resolution $\rm Mg_1$, Fe5709 and Fe5782
indices are bluer than those at low resolution, it raises the age
and metallicity. The discrepancy in these six indices is greater
for populations with $Z=0.03$ in comparison to $Z=0.01$.

At high resolution we compare the Lick/IDS spectral absorption
indices obtained by using the fitting functions with those
measured directly from the synthetic spectra, and see that Ca4455,
Fe4668, Mg$_{\rm b}$ and Na D indices obtained by the use of the
fitting functions are redder for all metallicities, Fe5709 is
redder at $Z=0.03$ and becomes to be bluer at $Z=0.01$ and 0.004,
and other indices are bluer for all metallicities than the
corresponding values measured directly from the synthetic spectra.
\end{abstract}


\section{Introduction}
In the previous papers \citep[hereinafter Paper I, II]{zha2004,
zha2005}, we took into account various known classes of binary
stars in evolutionary population synthesis (EPS) models, presented
integrated colours, integrated spectral energy distributions
(ISEDs) and 21 absorption-line indices [following the definitions
of \citealt{wor94}] defined by the Lick Observatory image
dissector scanner (referred to as the Lick/IDS) system for an
extensive set of instantaneous-burst binary stellar populations
(BSPs) with binary interactions, investigated the influences of
binary interactions and model input parameters on the results, and
found that the inclusion of binary interactions makes the
integrated $\rm {U-B}$, ${\rm B-V}$, ${\rm V-R}$ and ${\rm R-I}$
colours and the 21 Lick/IDS spectral indices substantially bluer.
In Papers I and II the corrected BaSeL-2.0 (i.e., non-calibrated
BaSeL-2.2) stellar spectra library of \citet*{lej97,lej98} was
used and low spectral resolution ISEDs ($10\, \rm \AA$ in the
ultraviolet and $20\, \rm \AA$ in the visible) were given.

In recent years, some spectroscopic galaxy surveys at intermediate
spectral resolution (such as the Sloan Digital Sky Survey, SDSS)
have been undertaken, and the data has been analyzed with EPS
models to understand galaxy formation and evolution. Therefore, a
EPS model that can predict the ISEDs at intermediate and high
spectral resolution is required. The high spectral resolution of
ISEDs can be used to predict the strength of numerous weak
absorption lines and the evolution of the profiles of the
strongest lines over a wide range of ages \citep{gon05}. And, the
higher the spectral resolution, the higher the constraining power
\citep{vaz99}. Therefore, in this paper we attempt to include high
resolution spectral information in our EPS models.

So far, the majority of EPS studies used low resolution stellar
spectral library as an ingredient. In the EPS models of
\citet{vaz99} and \citet{bru03} intermediate spectral resolution
ISEDs were provided: the former used a subsample of $\sim$ 500
stars from the original \citet{jon97} empirical stellar library
and presented the ISEDs of single stellar populations (SSPs) with
a spectral resolution of $\sim 1.8\, \rm \AA$ in two reduced
spectral regions: $3856-4476$ and $4795-5465\, \rm \AA$, the
latter used the STELIB library \citep{le03} and presented $3\, \rm
\AA$ resolution models covering $3200-9500\, \rm \AA$ with their
code GALAXEV. In addition, EPS models also can use the
intermediate-resolution empirical stellar library of
\citet{san03}, which comprises $\sim $ 1100 stars and covers the
spectral range $3500-7500$ \AA \,with a resolution of $\sim 2$
\AA.

With the emergence of new-generation empirical (e.g., the ELODIE
data base of \citealt{pru01}) and theoretical
\citep{ber03a,cas01,cha97,gon99,gon05,mun05,mur04,zwi02,zwi04}
stellar spectral libraries, EPS models can present the
high-resolution ISEDs and spectral absorption indices.
\citet{ber03b} used their library and explored the spectral
properties of SSPs in the optical (4880$-$5390\,\AA) and
ultraviolet regions (2275$-$2575\,\AA \,and 2690$-$2960\,\AA) at
$R=500\,000$ and $50\,000$, respectively. \citet{tan04} used the
library of \citet{mun05} and presented the Lick absorption-line
indices of SSPs from their 1-\AA\,resolution spectra. Recently,
\citet{gon05} used their library and presented high resolution
theoretical spectra ($\sim 0.3\, \rm \AA$) of SSPs in the
$3000-7000\, \rm \AA$ range.

In the few EPS studies to date that have presented intermediate
and high spectral resolution results, binary interactions are
neglected. Binary stars paly an important role in determining the
overall appearance of any realistic stellar population (
\citealt{xin05}, and Papers I and II for an extensive discussion),
therefore, in this paper we intend to use the high resolution
stellar spectral library in our EPS study of BSPs and present
their ISEDs and Lick/IDS absorption-feature indices. This set of
indices includes 21 indices of \citet{wor94} and 4 Balmer indices
defined by \citet{wor97}. In an instantaneous burst BSP all stars
are assumed to be born in binaries and born at the same time, and
$1 \times 10^6$ binary systems are comprised.

The outline of the paper is as follows: we describe our EPS models
and algorithm in Section 2; our results are presented in Section
3, and then finally, in Section 4, we give our conclusions.

\section{Model Description}

\subsection{Initialization of the binary population}

We first need to construct the instantaneous burst BSPs. A Monte
Carlo process is used to produce a population of $1 \times 10^6$
binary systems, and the initial state of each binary satisfies the
following input distributions:
\begin{enumerate}
\item the initial mass function (IMF) of the primaries, which
gives the relative number of the primaries in the mass range $m_1
\rightarrow m_1+ \rm dm_1$. In this study the initial mass of the
primary is chosen from the approximation to the IMF of
\citet{mil79} as given by \citet*[hereinafter EFT]{egg89},
\begin{equation}
m_1 = {\frac{ 0.19X }{(1-X)^{0.75}+0.032(1-X)^{0.25}}} \, ,
\label{mdis}
\end{equation}
where $X$ is a random variable uniformly distributed in the range
[0,1], and $m_1$ is the primary mass in units of ${\rm M_\odot}$.

\item the initial secondary-mass distribution, which is
assumed to be correlated with the initial primary-mass
distribution in this study. So, it depends on the initial
primary-mass (as set by equation \ref{mdis}) and the initial mass
ratio, $q$, distribution, which is assumed to be a uniform form
\citep[EFT 1989;][]{maz92,gol94},
\begin{equation}
n(q) = 1,  \ \ \ \ \     0 \leq q \leq 1,
\label{qdis}
\end{equation}
where $q = m_2/m_1$, $m_2$ is the secondary mass in units of ${\rm
M_\odot}$.

\item the distribution of orbital separations (or periods).
It is taken as constant in ${\rm log}\,a$ (where $a$ is the
separation) for wide binaries and falls off smoothly at close
separations:
\begin{equation}
a\,n(a) = \Bigl\{\matrix{\ a_{\rm sep}(a/a_0)^\psi , & \ \ \ a
\leq a_0, \cr a_{\rm sep}, \ \ \ \ \ \ \ \ \ & \ \ \ \ \ \ \ \ \ \
a_0 < a < a_1, \cr}
\label{adis}
\end{equation}
where $a_{\rm sep} \approx 0.070$, $a_0 = 10\,{\rm R_{\odot}}$,
$a_1=5.75\times 10^6 {\rm R_{\odot}}$ and $\psi \approx 1.2$.

\item the eccentricity distribution. A uniform form is
assumed:
\begin{equation}
e = X,
\label{edis}
\end{equation}
where $X$ is a random variable, as in Eq. (\ref{mdis}).
\end{enumerate}

After the initial state (the masses of the component stars,
$m_{\rm 1}$ and $m_{\rm 2}$, the separation $a$ and eccentricity
$e$ of the orbit) of a binary system in a BSP is set, we also need
to set the lower and upper mass cut-offs $m_{\rm l}$ and $m_{\rm
u}$ to the mass distributions and assign a metallicity $Z$ to the
stars: $m_{\rm l}$ and $m_{\rm u}$ are set as 0.1 and 100 ${\rm
M}_\odot$, respectively, $Z$ = 0.004, 0.01, 0.02, 0.03. The
relative age $\tau$ of the BSP is assigned within the range of
$1-15$\,Gyr.

\subsection{Input physics, parameters and algorithm}
\begin{figure}
\psfig{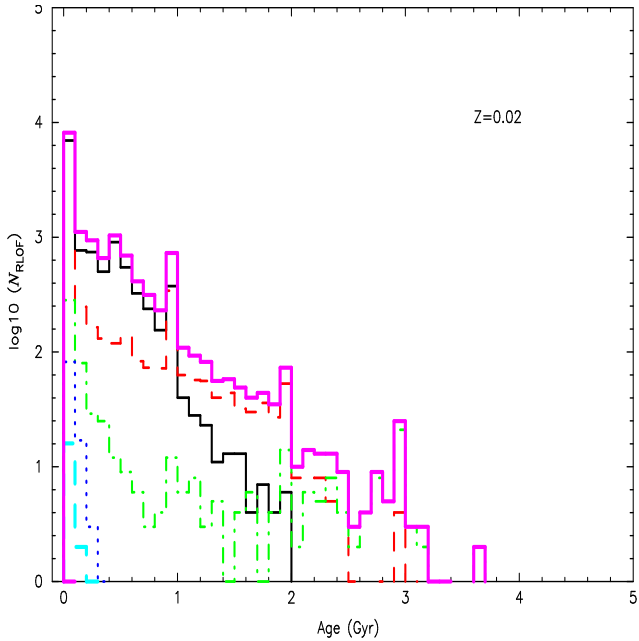}
\caption{Number of binary systems experiencing RLOF as a function
of time for solar-metallicity BSPs of $1 \times 10^6$ binaries.
Solid, dashed, dot-dashed, dotted, thick-dashed and thick-solid
lines represent the 1st, 2nd, 3rd, 4th, 5th and all RLOF,
respectively.} \label{nrlof-t}
\end{figure}

We use the rapid binary star evolution (BSE) algorithm of
\citet*{hur2002} to evolve each binary in the BSP to an age of
$\tau$, which gives us evolutionary parameters such as the stellar
luminosity $L$, effective temperature $T_{\rm eff}$, radius $R$,
current mass $m$ and the ratio of radius to Roche-lobe radius
$R/R_{\rm L}$ for the component stars. Next we use the high
resolution HRES stellar spectral library of \citet{gon05} to
transform the evolutionary parameters to stellar flux by
interpolating the flux grid in the log$T_{\rm eff}-$log$g-$[Fe/H]
plane, and then use equation (\ref{sp-lamda}) to obtain the
monochromatic flux for an instantaneous BSP of a particular age
and metallicity. For the spectral absorption-line indices in the
Lick/IDS system, we obtain them by two methods: (i) measure them
directly from the synthetic flux, (ii) use the empirical fitting
functions of \citet{wor94} and \citet{wor97} to assign the
Lick/IDS absorption indices to a star with a set of given
evolutionary parameters, and then use equations
(\ref{inte-EW})$-$(\ref{inte-mag}) to derive the indices for a
BSP.

The integrated monochromatic flux of a BSP is defined as
\begin{equation}
F_{\lambda,\tau,Z} = \sum_{{\rm k=1}}^{{\rm n}} \ f_{\lambda},
\label{sp-lamda}
\end{equation}
where $f_{\lambda}$ is the SED of the $k$th star.

The integrated absorption feature index of the Lick/IDS system is
a flux-weighted one. For the $i$th atomic absorption line, it is
expressed in equivalent width ($W$, in angstroms),
\begin{equation}
W_{i,\tau,Z} = {\frac{\sum_{{\rm k=1}}^{{\rm n}} \ w_{{\rm i}} \
\cdot f_{i,{\rm C}\lambda} }{\sum_{{\rm k=1}}^{{\rm n}} \
f_{i,{\rm C}\lambda}}},
\label{inte-EW}
\end{equation}
where $w_{{\rm i}}$ is the equivalent width of the $i$th index of
the $k$th star, and $ f_{i,{\rm C} \lambda}$ is the continuum flux
at the midpoint of the $i$th `feature' passband. The local
continuum for the $i$-th index is the run of flux defined by
drawing a straight line from the midpoint of the blue
pseudocontinuum level to the midpoint of the red pseudocontinuum
level, the flux at the midpoint of pseudocontinuum is a average
one and obtained by
\begin{equation}
f_{i,p}={{{\int_{\lambda_{i,1}}^{\lambda_{i,2}} f_{\lambda} \rm{d}
\lambda}} \over {\lambda_{i,2} - \lambda_{i,1}}},
\label{fint}
\end{equation}
where $f_\lambda$ is the stellar flux, as in Eq. (\ref{sp-lamda}),
$\lambda_{i,1}$ and $\lambda_{i,2}$ are the wavelength limits of
the $i$-th pseudocontinuum sideband. For the $i$th molecular line,
the feature index is expressed in magnitude,
\begin{equation}
C_{i,\tau,Z} = -2.5 \log {\frac{\sum_{{\rm k=1}}^{{\rm n}} \
10^{-0.4 c_{{\rm i}}} \cdot \ f_{i,{\rm C}\lambda} }{\sum_{{\rm
k=1}}^{{\rm n}} \ f_{i,{\rm C}\lambda}}},
\label{inte-mag}
\end{equation}
where $c_{{\rm i}}$ is the magnitude of the $i$th index of the
$k$th star.

Detailed descriptions of the BSE package of \citet{hur2002} and
the empirical fitting functions of \citet{wor94} have been
presented in Papers I and II, here we only mention several
important input parameters required in the BSE code: the
efficiency of common envelope ejection $\alpha_{\rm CE}$ is taken
as 1.0, the Reimers wind mass-loss coefficient $\eta$ is set
constant at 0.3, and the tidal enhancement parameter $B=0.0$.

Following, we present a brief description of the HRES stellar
spectral library of \citet{gon05}. A full discussion is given in
\cite{mar05}. This library includes the synthetic stellar spectra
from 3000 to $7000\, \rm \AA$ with a final spectral sampling of
$0.3\, \rm \AA$. The spectra span a range of effective temperature
from 3000 to 55000K, with variable steps from 500 to 2500K, and a
surface gravity ${\rm log}g = -0.5$ to 5.5 with dex steps of 0.25
and 0.5. For each temperature, the minimum gravity is set by the
Eddington limit. The library covers several metallicities: twice
solar, solar, half solar and 1/10 solar. Solar abundance ratios
for all the elements, and a helium abundance of He/H = 0.1 by
number are assumed. This high spectral resolution stellar library
is based on Kurucz local thermodynamic equilibrium (LTE)
atmospheres \citep{kur93} and the program SYNSPEC \citep*{hub95}
for stars with $8000 \le T_{\rm eff} \le 27000$ K, the program
SPECTRUM \citep{gra94} and Kurucz atmospheres for stars with $4750
\le T_{\rm eff} \le 7750$ K, non-LTE line-blanketed models for hot
($27500 \le T_{\rm eff} \le 55000$ K, \citealt{lan03}), and
PHOENIX LTE line-blanketed models for cool stars ($3000 \le T_{\rm
eff} \le 4500$ K, \citealt{hau99,all01}). This library is composed
of 1650 spectra.

By adopting the above set of input distributions and parameters,
in Fig. \ref{nrlof-t} we present the number of binary systems
experiencing Roche lobe overflow (RLOF) as a function of time for
solar-metallicity BSPs of $1 \times 10^6$ binaries. Because some
binary systems would experience RLOF for several times, in Fig.
\ref{nrlof-t} solid, dashed, dot-dashed, dotted and thick-dashed
and thick-solid lines represent the 1st, 2nd, 3rd, 4th, 5th and
all RLOF, respectively. From Fig. \ref{nrlof-t} we see that RLOF
mainly happens at early age, and the number decreases with time.
During the past 15\,Gyr $\sim$ 11.6 per cent of the binaries would
experience RLOF.

\section{Results}

\subsection{The integrated spectral energy distribution}
\begin{figure}
\psfig{file=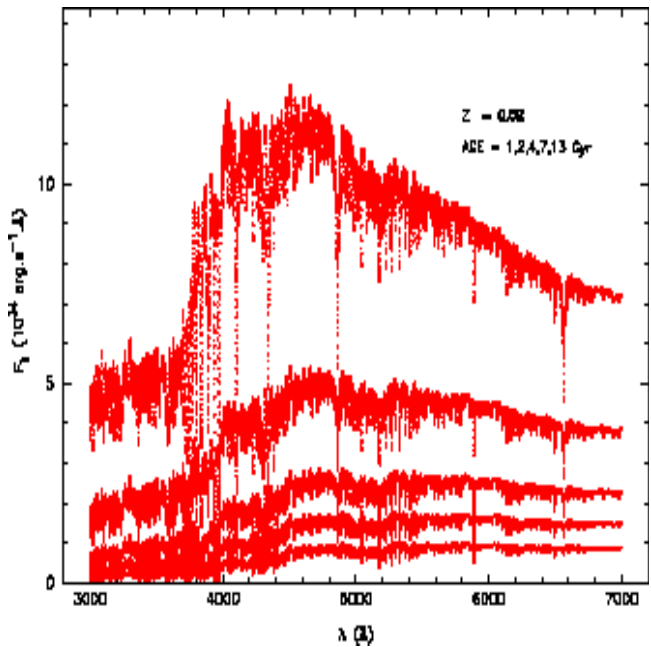,height=8.5cm,width=8.5cm,bbllx=503pt,bblly=0pt,bburx=0pt,bbury=665pt,clip=,angle=270}
\caption{The high spectral resolution integrated spectral energy
distributions as a function of age for solar-metallicity BSPs.
From top to bottom, the ages $\tau = 1, 2, 4, 7, 13$\,Gyr,
respectively.} \label{ised-hs-t}
\end{figure}

\begin{figure*}
\centerline{
\psfig{file=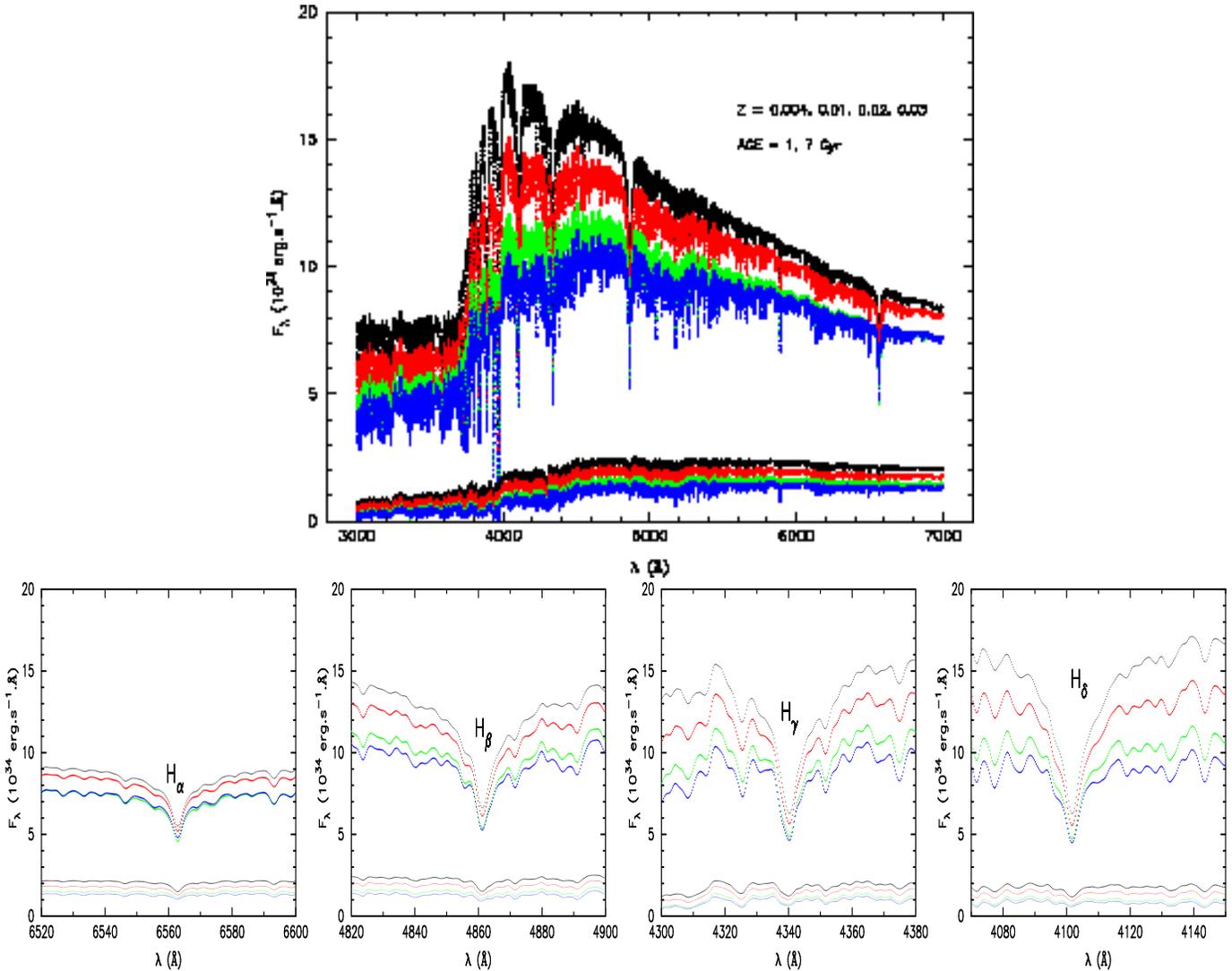,height=8.5cm,width=10.5cm,bbllx=508pt,bblly=0pt,bburx=0pt,bbury=667pt,clip=,angle=270}}
\centerline{
\psfig{file=MF414rv2fig32.ps,height=5.8cm,width=4.5cm,bbllx=532pt,bblly=103pt,bburx=128pt,bbury=646pt,clip=,angle=270}
\psfig{file=MF414rv2fig33.ps,height=5.8cm,width=4.5cm,bbllx=532pt,bblly=103pt,bburx=128pt,bbury=646pt,clip=,angle=270}
\psfig{file=MF414rv2fig34.ps,height=5.8cm,width=4.5cm,bbllx=532pt,bblly=103pt,bburx=128pt,bbury=646pt,clip=,angle=270}
\psfig{file=MF414rv2fig35.ps,height=5.8cm,width=4.5cm,bbllx=532pt,bblly=103pt,bburx=128pt,bbury=646pt,clip=,angle=270}}
\caption{The high resolution integrated spectral energy
distributions (top panel) and Balmer lines (bottom panel) as a
function of metallicity (From top to bottom, the metallicity $Z =
0.004, 0.01, 0.02$ and 0.03, respectively) for BSPs at ages $\tau=
1$ and 7 Gyr.} \label{ised-hs-z}
\end{figure*}

\begin{figure}
\psfig{file=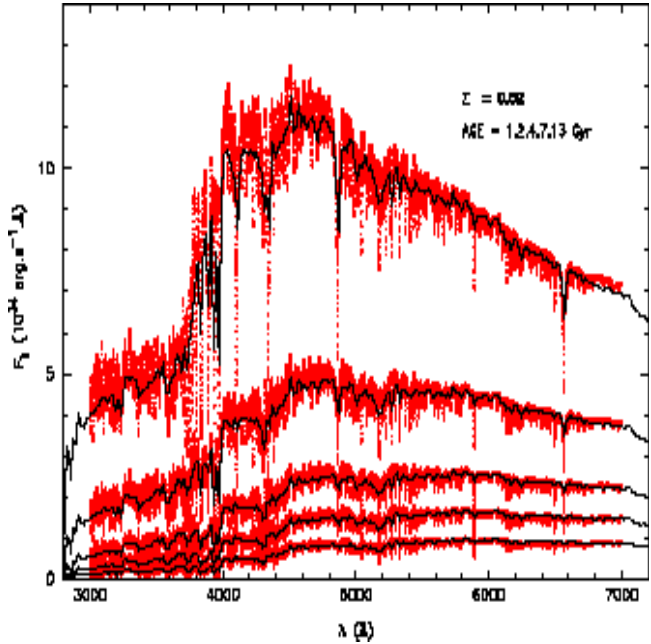,height=8.5cm,width=8.5cm,bbllx=503pt,bblly=0pt,bburx=0pt,bbury=665pt,clip=,angle=270}
\caption{Comparison of the high resolution (solid line) with low
resolution (dots) integrated spectral energy distributions for
solar-metallicity BSPs at ages $\tau = 1, 2, 4, 7$ and 13 \,Gyr
(from top to bottom, respectively).} \label{ised-com}
\end{figure}

\begin{figure}
\psfig{file=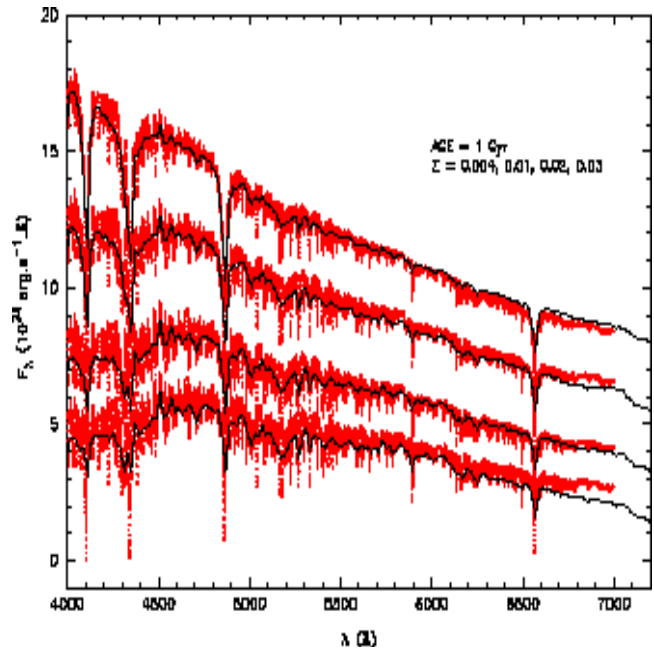,height=8.5cm,width=8.5cm,bbllx=0pt,bblly=0pt,bburx=505pt,bbury=667pt,clip=,angle=270}
\caption{Comparison between the high (solid line) and low
resolution (dots) integrated spectral energy distributions for
$\tau = 1$\,Gyr BSPs with $Z = 0.004, 0.01, 0.02$ and 0.03 (from
top to bottom, respectively). For the sake of clarity the fluxes
at metallicities of 0.01, 0.02 and 0.03 are shifted downwards by
an amount of 1.5.} \label{ised-com2}
\end{figure}

In this part we present ISEDs covering the range $3000 - 7000\,
\rm \AA$ with a resolution of $0.3\, \rm \AA$ for instantaneous
burst BSPs with binary interactions over a large range of age and
metallicity: $1 \leq \tau \leq 15$\,Gyr and $-1.3 \leq {\rm
[Fe/H]} \leq +0.2$. For each model, a total of 1.0$\times 10^6$
binaries are evolved according to the algorithm given in the
previous section. The full set ISEDs are available on request from
the authors.

In Fig. \ref{ised-hs-t} we give the high spectral resolution ISED
evolution for solar-metallicity BSPs at ages $\tau = 1, 2, 4, 7$
and 13 \,Gyr in the wavelength range of $3000 - 7000\, \rm \AA$.
It shows that the continuum tends to be redder with increasing the
age of the BSP, the variation in the shape of the continuum with
time is significant at early age, and is almost constant at
intermediate and late ages. Additionally, Fig. \ref{ised-hs-t}
shows that the strength of the Balmer lines decreases with the age
of the BSP.

The metallicity effect on ISED evolution of BSPs is shown in the
top panel of Fig. \ref{ised-hs-z}, which presents the ISEDs of
young ($\tau =1$\,Gyr) and intermediate-age ($\tau =7$\,Gyr) BSPs
at four different metallicities: $Z = 0.004, 0.01, 0.02, 0.03$.
The metallicity effect is also exhibited in the slope of the
continuum and in the strength of the metallic lines. The continuum
tends to be redder with increasing metallicity and metallic lines
to be stronger. The bottom panel of Fig. \ref{ised-hs-z} gives the
evolution of Bamler lines with metallicity, it shows that their
strength tends to be weaker when increasing metallicity.

In Fig. \ref{ised-com} we compare the ISEDs at high and low
spectral resolution ($10\, \rm \AA$ in the ultraviolet and $20\,
\rm \AA$ in the visible) for solar-metallicity BSPs at ages $\tau
= 1, 2 , 4, 7$ and 13 \,Gyr. These low resolution ISEDs are from
Model A of Paper II, which includes binary interactions and adopts
the same input parameters and distributions as this paper except
for using the corrected BaSeL-2.0 (i.e., non-calibrated BaSeL-2.2)
stellar spectral library of \citet{lej97,lej98}. Fig.
\ref{ised-com} shows that there is a good agreement in the
continuum for solar-metallicity BSPs between two studies.
Comparison of the continuum for BSPs with non-solar metallicity
shows that there is a difference existed {\rm at longer}
wavelengths for BSPs at ages $\tau = 1$ and 2\,Gyr. In Fig
\ref{ised-com2} we give the ISEDs at high and low resolution for
$\tau = 1$\,Gyr BSP with four metallicities. It shows that at
longer wavelengths the high resolution continuum is lower (redder)
for BSPs with $Z=0.004$, while greater (bluer) for BSPs with
$Z=0.01$ and 0.03 than that at low resolution.

\subsection{Lick spectral absorption feature indices}
\begin{figure}
\psfig{file=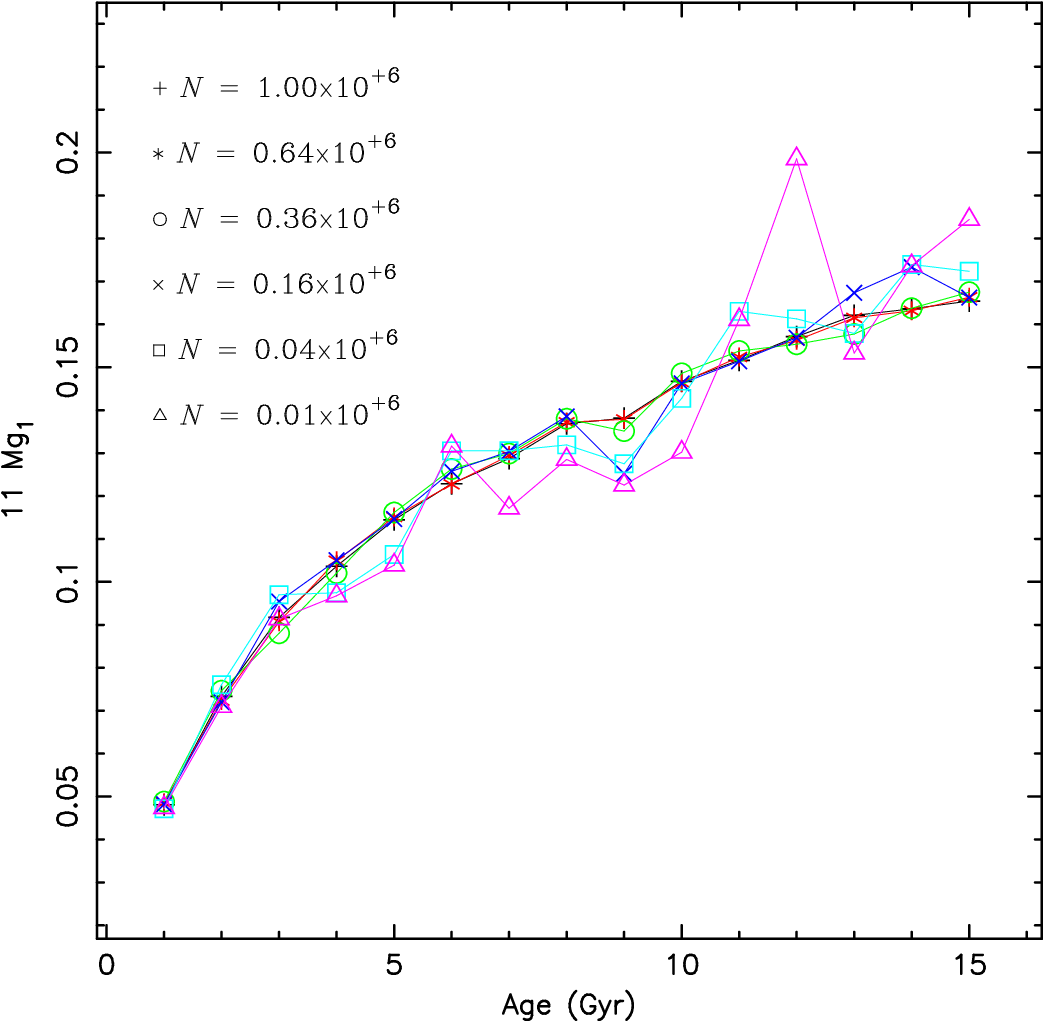,height=8.5cm,width=8.5cm,bbllx=572pt,bblly=167pt,bburx=80pt,bbury=672pt,clip=,angle=270}
\caption{Sensibility of Mg$_1$ index to the number of binary
systems in the BSPs.} \label{mc}
\end{figure}

At high spectral resolution we use two methods to obtain 25
Lick/IDS spectral absorption indices defined by \citet{wor94} and
\citet{wor97}: (1) directly compute them from the high-resolution
synthetic spectra, (2) obtain them by using the empirical fitting
functions of \citet{wor94} and \citet{wor97}.

All the following results are for BSPs of $10^6$ binaries. To
verify that our results are stable for $10^6$ binaries, we perform
several simulations, the number of binaries in each simulation is
$10^4, 4 \times 10^4, 1.6 \times 10^5, 3.6 \times 10^5$ and $6.4
\times 10^5$, respectively. As an example, in Fig. \ref{mc} we
plot the evolution of Mg$_1$ index as a function of the number of
binaries in the simulation, we see that the Mg$_1$ index
fluctuates around the value for $10^6$ binaries when the number of
binaries is not so enough, with the increase of the number of
binaries the fluctuation would decrease. For $10^4$ binaries the
discrepancy reaches to $\sim 0.05\,$\AA \, at $\tau=12\,$Gyr, for
$3.6 \times 10^5$ binaries Mg$_1$ agrees with the value for $10^6$
binaries within a small range, and for $6.4 \times 10^5$ binaries
the discrepancy is insignificant. Therefore the results for $10^6$
binaries in the simulation are stable.

\subsubsection{Results by the use of the fitting functions}
\begin{table*}
\tiny
\centering
\begin{minipage}{200mm}
\caption{Lick/IDS spectral absorption feature indices of BSPs,
derived by using the empirical fitting functions (Section 3.2.1).}
\begin{tabular}{rrrrrrrrrrrrrrrr}
\hline
  \multicolumn{16}{c}{Age (Gyr)} \\
  &   1.00  &   2.00  &   3.00  &   4.00  &   5.00  &   6.00  &   7.00  &   8.00  &   9.00  &  10.00  &  11.00  &  12.00  &  13.00  &  14.00  &  15.00  \\
\hline
\multicolumn{16}{c}{$Z$ = 0.004} \\
$\rm CN_1$   &  -0.184  &  -0.136  &  -0.106  &  -0.079  &  -0.065  &  -0.051  &  -0.048  &  -0.051  &  -0.038  &  -0.034  &  -0.032  &  -0.044  &  -0.038  &  -0.035  &  -0.036  \\
$\rm CN_2$   &  -0.110  &  -0.077  &  -0.056  &  -0.037  &  -0.027  &  -0.015  &  -0.013  &  -0.015  &  -0.006  &  -0.003  &  -0.002  &  -0.011  &  -0.006  &  -0.005  &  -0.006  \\
Ca4227       &   0.157  &   0.334  &   0.436  &   0.523  &   0.589  &   0.657  &   0.699  &   0.683  &   0.709  &   0.744  &   0.777  &   0.770  &   0.774  &   0.830  &   0.802  \\
G4300        &  -0.883  &   0.715  &   1.730  &   2.585  &   3.081  &   3.504  &   3.652  &   3.585  &   4.011  &   4.213  &   4.309  &   3.987  &   4.200  &   4.366  &   4.360  \\
Fe4383       &  -0.377  &   0.419  &   1.131  &   1.703  &   2.026  &   2.341  &   2.433  &   2.501  &   2.691  &   2.824  &   2.887  &   2.848  &   3.021  &   3.130  &   3.169  \\
Ca4455       &   0.213  &   0.480  &   0.639  &   0.757  &   0.845  &   0.919  &   0.961  &   0.951  &   0.982  &   1.015  &   1.049  &   1.041  &   1.055  &   1.085  &   1.067  \\
Fe4531       &   1.041  &   1.556  &   1.830  &   2.054  &   2.187  &   2.320  &   2.369  &   2.340  &   2.422  &   2.481  &   2.521  &   2.493  &   2.547  &   2.580  &   2.562  \\
Fe4668       &  -0.122  &   0.377  &   0.816  &   1.006  &   1.275  &   1.413  &   1.595  &   1.604  &   1.448  &   1.451  &   1.599  &   1.708  &   1.626  &   1.705  &   1.590  \\
$\rm H_\beta$&   5.564  &   4.352  &   3.585  &   2.984  &   2.690  &   2.415  &   2.360  &   2.394  &   2.159  &   2.054  &   2.018  &   2.179  &   2.030  &   1.956  &   1.950  \\
Fe5015       &   2.028  &   2.872  &   3.299  &   3.412  &   3.700  &   3.796  &   4.025  &   4.039  &   3.793  &   3.832  &   3.998  &   4.137  &   3.897  &   3.989  &   3.820  \\
$\rm Mg_1 $  &   0.012  &   0.014  &   0.019  &   0.025  &   0.027  &   0.032  &   0.033  &   0.033  &   0.037  &   0.040  &   0.042  &   0.042  &   0.047  &   0.048  &   0.049  \\
$\rm Mg_2$   &   0.062  &   0.078  &   0.092  &   0.100  &   0.109  &   0.116  &   0.121  &   0.122  &   0.122  &   0.127  &   0.133  &   0.138  &   0.144  &   0.144  &   0.142  \\
$\rm Mg_b$   &   1.155  &   1.486  &   1.745  &   1.808  &   2.059  &   2.039  &   2.187  &   2.224  &   2.079  &   2.132  &   2.267  &   2.431  &   2.480  &   2.480  &   2.413  \\
Fe5270       &   0.812  &   1.229  &   1.476  &   1.641  &   1.764  &   1.869  &   1.932  &   1.916  &   1.946  &   1.993  &   2.035  &   2.050  &   2.068  &   2.107  &   2.085  \\
Fe5335       &   0.635  &   0.917  &   1.114  &   1.244  &   1.349  &   1.438  &   1.496  &   1.496  &   1.502  &   1.543  &   1.590  &   1.625  &   1.654  &   1.682  &   1.661  \\
Fe5406       &   0.293  &   0.492  &   0.618  &   0.722  &   0.784  &   0.865  &   0.881  &   0.879  &   0.914  &   0.942  &   0.967  &   0.971  &   1.013  &   1.022  &   1.008  \\
Fe5709       &   0.320  &   0.450  &   0.508  &   0.577  &   0.590  &   0.638  &   0.622  &   0.607  &   0.665  &   0.678  &   0.672  &   0.644  &   0.679  &   0.678  &   0.681  \\
Fe5782       &   0.208  &   0.294  &   0.345  &   0.400  &   0.415  &   0.455  &   0.450  &   0.440  &   0.469  &   0.482  &   0.481  &   0.470  &   0.501  &   0.496  &   0.495  \\
Na D         &   1.193  &   1.197  &   1.306  &   1.363  &   1.455  &   1.500  &   1.592  &   1.620  &   1.572  &   1.660  &   1.685  &   1.784  &   1.814  &   1.858  &   1.870  \\
$\rm TiO_1$  &   0.033  &   0.033  &   0.035  &   0.030  &   0.034  &   0.032  &   0.039  &   0.040  &   0.029  &   0.030  &   0.034  &   0.041  &   0.033  &   0.034  &   0.030  \\
$\rm TiO_2$  &   0.034  &   0.035  &   0.040  &   0.032  &   0.040  &   0.037  &   0.049  &   0.052  &   0.032  &   0.033  &   0.042  &   0.053  &   0.040  &   0.042  &   0.035  \\
$\rm H\delta_A$  &   8.599  &   6.626  &   4.769  &   3.260  &   2.467  &   1.824  &   1.574  &   1.517  &   0.959  &   0.678  &   0.536  &   1.062  &   0.679  &   0.347  &   0.349  \\
$\rm H\gamma_A$  &   7.834  &   4.965  &   2.626  &   0.815  &  -0.215  &  -1.137  &  -1.513  &  -1.587  &  -2.383  &  -2.748  &  -2.975  &  -2.340  &  -2.734  &  -3.164  &  -3.137  \\
$\rm H\delta_F$  &   6.071  &   4.772  &   3.681  &   2.756  &   2.228  &   1.865  &   1.720  &   1.698  &   1.384  &   1.218  &   1.148  &   1.481  &   1.246  &   1.088  &   1.101  \\
$\rm H\gamma_F$  &   6.032  &   4.429  &   3.148  &   2.108  &   1.491  &   0.994  &   0.785  &   0.765  &   0.317  &   0.116  &  -0.012  &   0.321  &   0.112  &  -0.141  &  -0.128  \\

\multicolumn{16}{c}{$Z$ = 0.01} \\
$\rm CN_1$   &  -0.155  &  -0.088  &  -0.057  &  -0.046  &  -0.036  &  -0.022  &  -0.023  &  -0.013  &  -0.006  &  -0.006  &  -0.004  &  -0.003  &   0.000  &   0.005  &   0.004  \\
$\rm CN_2$   &  -0.092  &  -0.042  &  -0.018  &  -0.009  &  -0.001  &   0.012  &   0.012  &   0.021  &   0.027  &   0.026  &   0.028  &   0.029  &   0.031  &   0.034  &   0.034  \\
Ca4227       &   0.309  &   0.563  &   0.689  &   0.754  &   0.767  &   0.892  &   0.856  &   0.931  &   0.984  &   0.993  &   1.010  &   1.033  &   1.037  &   1.050  &   1.055  \\
G4300        &   0.134  &   2.241  &   3.195  &   3.526  &   3.806  &   4.208  &   4.156  &   4.441  &   4.657  &   4.694  &   4.794  &   4.809  &   4.902  &   5.114  &   5.072  \\
Fe4383       &   0.421  &   1.820  &   2.592  &   2.910  &   3.318  &   3.565  &   3.625  &   3.851  &   4.026  &   4.088  &   4.137  &   4.203  &   4.304  &   4.407  &   4.421  \\
Ca4455       &   0.497  &   0.862  &   1.024  &   1.105  &   1.155  &   1.253  &   1.237  &   1.306  &   1.356  &   1.363  &   1.382  &   1.402  &   1.421  &   1.439  &   1.453  \\
Fe4531       &   1.526  &   2.166  &   2.443  &   2.560  &   2.660  &   2.798  &   2.782  &   2.898  &   2.972  &   2.987  &   3.000  &   3.028  &   3.066  &   3.089  &   3.111  \\
Fe4668       &   0.837  &   2.010  &   2.460  &   2.752  &   2.830  &   3.178  &   3.081  &   3.246  &   3.381  &   3.360  &   3.421  &   3.477  &   3.480  &   3.468  &   3.485  \\
$\rm H_\beta$&   5.158  &   3.527  &   2.873  &   2.666  &   2.503  &   2.298  &   2.305  &   2.135  &   2.024  &   1.998  &   1.961  &   1.935  &   1.882  &   1.787  &   1.785  \\
Fe5015       &   3.063  &   4.022  &   4.264  &   4.531  &   4.476  &   4.794  &   4.646  &   4.746  &   4.864  &   4.815  &   4.923  &   5.010  &   4.976  &   4.950  &   4.991  \\
$\rm Mg_1$   &   0.012  &   0.024  &   0.035  &   0.039  &   0.046  &   0.051  &   0.052  &   0.059  &   0.062  &   0.064  &   0.064  &   0.066  &   0.069  &   0.070  &   0.073  \\
$\rm Mg_2$   &   0.080  &   0.115  &   0.131  &   0.142  &   0.149  &   0.163  &   0.162  &   0.171  &   0.179  &   0.182  &   0.184  &   0.188  &   0.192  &   0.192  &   0.196  \\
$\rm Mg_b$   &   1.505  &   2.137  &   2.263  &   2.482  &   2.479  &   2.752  &   2.679  &   2.770  &   2.893  &   2.920  &   3.021  &   3.101  &   3.090  &   3.102  &   3.113  \\
Fe5270       &   1.305  &   1.846  &   2.071  &   2.185  &   2.258  &   2.380  &   2.371  &   2.459  &   2.521  &   2.536  &   2.558  &   2.587  &   2.610  &   2.625  &   2.650  \\
Fe5335       &   1.015  &   1.496  &   1.702  &   1.805  &   1.878  &   1.993  &   1.978  &   2.068  &   2.128  &   2.144  &   2.155  &   2.186  &   2.209  &   2.213  &   2.243  \\
Fe5406       &   0.513  &   0.837  &   1.007  &   1.070  &   1.145  &   1.218  &   1.224  &   1.297  &   1.341  &   1.346  &   1.348  &   1.371  &   1.392  &   1.394  &   1.416  \\
Fe5709       &   0.476  &   0.613  &   0.704  &   0.720  &   0.768  &   0.779  &   0.798  &   0.828  &   0.839  &   0.841  &   0.834  &   0.838  &   0.852  &   0.857  &   0.865  \\
Fe5782       &   0.323  &   0.463  &   0.539  &   0.560  &   0.599  &   0.621  &   0.628  &   0.660  &   0.674  &   0.674  &   0.662  &   0.669  &   0.681  &   0.675  &   0.684  \\
Na D         &   1.323  &   1.677  &   1.834  &   1.948  &   2.006  &   2.145  &   2.144  &   2.219  &   2.295  &   2.351  &   2.380  &   2.417  &   2.441  &   2.465  &   2.501  \\
$\rm TiO_1$  &   0.036  &   0.040  &   0.037  &   0.041  &   0.037  &   0.042  &   0.038  &   0.038  &   0.040  &   0.039  &   0.040  &   0.041  &   0.040  &   0.039  &   0.040  \\
$\rm TiO_2$  &   0.037  &   0.047  &   0.044  &   0.051  &   0.045  &   0.057  &   0.047  &   0.049  &   0.053  &   0.051  &   0.053  &   0.056  &   0.054  &   0.051  &   0.053  \\
$\rm H\delta_A$  &   7.873  &   3.990  &   2.279  &   1.729  &   1.288  &   0.481  &   0.526  &   0.020  &  -0.448  &  -0.470  &  -0.687  &  -0.761  &  -0.876  &  -1.333  &  -1.229  \\
$\rm H\gamma_A$  &   6.677  &   1.902  &  -0.309  &  -1.093  &  -1.710  &  -2.707  &  -2.638  &  -3.291  &  -3.859  &  -3.901  &  -4.149  &  -4.264  &  -4.375  &  -4.905  &  -4.806  \\
$\rm H\delta_F$  &   5.447  &   3.210  &   2.226  &   1.925  &   1.690  &   1.275  &   1.342  &   1.074  &   0.830  &   0.833  &   0.728  &   0.691  &   0.634  &   0.412  &   0.464  \\
$\rm H\gamma_F$  &   5.315  &   2.758  &   1.541  &   1.104  &   0.781  &   0.236  &   0.274  &  -0.072  &  -0.372  &  -0.408  &  -0.557  &  -0.620  &  -0.676  &  -0.969  &  -0.913  \\

\multicolumn{16}{c}{$Z$ = 0.02} \\
$\rm CN_1$   &  -0.121  &  -0.060  &  -0.028  &  -0.009  &   0.006  &   0.018  &   0.030  &   0.023  &   0.035  &   0.048  &   0.050  &   0.045  &   0.058  &   0.032  &   0.065  \\
$\rm CN_2$   &  -0.065  &  -0.018  &   0.009  &   0.026  &   0.040  &   0.050  &   0.062  &   0.057  &   0.068  &   0.080  &   0.083  &   0.077  &   0.091  &   0.070  &   0.097  \\
Ca4227       &   0.428  &   0.714  &   0.878  &   0.964  &   1.045  &   1.085  &   1.187  &   1.155  &   1.206  &   1.310  &   1.329  &   1.251  &   1.356  &   1.256  &   1.434  \\
G4300        &   1.045  &   2.947  &   3.784  &   4.210  &   4.583  &   4.873  &   5.121  &   4.969  &   5.283  &   5.526  &   5.536  &   5.459  &   5.692  &   5.051  &   5.835  \\
Fe4383       &   1.376  &   2.826  &   3.718  &   4.195  &   4.533  &   4.828  &   5.101  &   5.135  &   5.346  &   5.564  &   5.647  &   5.690  &   5.838  &   5.595  &   6.059  \\
Ca4455       &   0.761  &   1.107  &   1.294  &   1.395  &   1.481  &   1.546  &   1.623  &   1.604  &   1.670  &   1.746  &   1.761  &   1.732  &   1.800  &   1.715  &   1.856  \\
Fe4531       &   1.956  &   2.540  &   2.833  &   3.001  &   3.129  &   3.231  &   3.343  &   3.327  &   3.418  &   3.525  &   3.547  &   3.522  &   3.613  &   3.487  &   3.704  \\
Fe4668       &   2.016  &   3.171  &   3.869  &   4.212  &   4.513  &   4.733  &   5.006  &   4.911  &   5.171  &   5.415  &   5.475  &   5.288  &   5.477  &   5.340  &   5.652  \\
$\rm H_\beta$&   4.660  &   3.156  &   2.642  &   2.386  &   2.197  &   2.050  &   1.910  &   1.967  &   1.802  &   1.697  &   1.654  &   1.681  &   1.563  &   1.800  &   1.467  \\
Fe5015       &   3.587  &   4.375  &   4.833  &   4.993  &   5.179  &   5.276  &   5.453  &   5.344  &   5.509  &   5.697  &   5.712  &   5.565  &   5.741  &   5.604  &   5.782  \\
$\rm Mg_1$   &   0.021  &   0.042  &   0.056  &   0.066  &   0.071  &   0.077  &   0.084  &   0.086  &   0.090  &   0.094  &   0.097  &   0.097  &   0.101  &   0.099  &   0.108  \\
$\rm Mg_2$   &   0.102  &   0.141  &   0.168  &   0.184  &   0.196  &   0.208  &   0.219  &   0.220  &   0.231  &   0.240  &   0.245  &   0.240  &   0.249  &   0.246  &   0.260  \\
$\rm Mg_b$   &   1.696  &   2.229  &   2.654  &   2.858  &   3.053  &   3.268  &   3.374  &   3.345  &   3.644  &   3.656  &   3.705  &   3.621  &   3.696  &   3.687  &   3.848  \\
Fe5270       &   1.714  &   2.207  &   2.471  &   2.614  &   2.729  &   2.806  &   2.905  &   2.890  &   2.971  &   3.057  &   3.088  &   3.059  &   3.157  &   3.070  &   3.206  \\
Fe5335       &   1.454  &   1.935  &   2.187  &   2.325  &   2.426  &   2.504  &   2.600  &   2.593  &   2.671  &   2.746  &   2.773  &   2.745  &   2.817  &   2.763  &   2.887  \\
Fe5406       &   0.804  &   1.144  &   1.330  &   1.449  &   1.519  &   1.582  &   1.650  &   1.656  &   1.708  &   1.758  &   1.783  &   1.771  &   1.829  &   1.776  &   1.881  \\
Fe5709       &   0.658  &   0.809  &   0.882  &   0.936  &   0.962  &   0.985  &   1.006  &   1.008  &   1.016  &   1.031  &   1.037  &   1.039  &   1.061  &   1.025  &   1.069  \\
Fe5782       &   0.490  &   0.632  &   0.704  &   0.757  &   0.782  &   0.805  &   0.830  &   0.832  &   0.846  &   0.863  &   0.871  &   0.864  &   0.889  &   0.862  &   0.901  \\
Na D         &   1.635  &   2.123  &   2.397  &   2.560  &   2.689  &   2.777  &   2.901  &   2.926  &   3.024  &   3.121  &   3.182  &   3.176  &   3.302  &   3.262  &   3.400  \\
$\rm TiO_1$  &   0.026  &   0.029  &   0.033  &   0.033  &   0.035  &   0.035  &   0.038  &   0.036  &   0.040  &   0.042  &   0.042  &   0.038  &   0.040  &   0.042  &   0.041  \\
$\rm TiO_2$  &   0.021  &   0.030  &   0.040  &   0.041  &   0.044  &   0.046  &   0.051  &   0.049  &   0.056  &   0.059  &   0.060  &   0.052  &   0.056  &   0.059  &   0.059  \\
$\rm H\delta_A$  &   6.757  &   2.674  &   1.211  &   0.400  &  -0.372  &  -0.988  &  -1.552  &  -1.242  &  -2.148  &  -2.532  &  -2.798  &  -2.385  &  -3.220  &  -2.042  &  -3.449  \\
$\rm H\gamma_A$  &   5.213  &   0.371  &  -1.653  &  -2.697  &  -3.645  &  -4.361  &  -5.012  &  -4.644  &  -5.673  &  -6.065  &  -6.362  &  -5.824  &  -6.796  &  -5.488  &  -6.994  \\
$\rm H\delta_F$  &   4.713  &   2.545  &   1.744  &   1.343  &   0.963  &   0.654  &   0.409  &   0.588  &   0.121  &  -0.035  &  -0.142  &   0.049  &  -0.318  &   0.311  &  -0.396  \\
$\rm H\gamma_F$  &   4.502  &   1.993  &   0.892  &   0.329  &  -0.182  &  -0.554  &  -0.910  &  -0.733  &  -1.237  &  -1.472  &  -1.620  &  -1.361  &  -1.843  &  -1.194  &  -1.985  \\

\multicolumn{16}{c}{$Z$ = 0.03} \\
$\rm CN_1$   &  -0.102  &  -0.040  &  -0.017  &   0.003  &   0.018  &   0.029  &   0.034  &   0.044  &   0.057  &   0.058  &   0.057  &   0.076  &   0.092  &   0.087  &   0.084  \\
$\rm CN_2$   &  -0.050  &  -0.001  &   0.020  &   0.038  &   0.052  &   0.063  &   0.068  &   0.078  &   0.090  &   0.092  &   0.093  &   0.111  &   0.128  &   0.122  &   0.119  \\
Ca4227       &   0.575  &   0.881  &   1.058  &   1.213  &   1.252  &   1.350  &   1.377  &   1.496  &   1.603  &   1.543  &   1.567  &   1.657  &   1.789  &   1.827  &   1.723  \\
G4300        &   1.595  &   3.521  &   4.031  &   4.490  &   4.835  &   4.990  &   5.105  &   5.240  &   5.530  &   5.518  &   5.364  &   5.762  &   5.921  &   5.923  &   5.813  \\
Fe4383       &   1.856  &   3.542  &   4.299  &   4.710  &   5.123  &   5.401  &   5.613  &   5.798  &   5.990  &   6.135  &   6.228  &   6.426  &   6.725  &   6.655  &   6.616  \\
Ca4455       &   0.967  &   1.318  &   1.486  &   1.609  &   1.669  &   1.736  &   1.764  &   1.846  &   1.931  &   1.922  &   1.935  &   2.014  &   2.084  &   2.102  &   2.077  \\
Fe4531       &   2.221  &   2.796  &   3.039  &   3.226  &   3.356  &   3.454  &   3.520  &   3.618  &   3.740  &   3.746  &   3.786  &   3.891  &   3.999  &   3.996  &   3.986  \\
Fe4668       &   3.107  &   4.282  &   5.019  &   5.480  &   5.567  &   5.870  &   5.897  &   6.320  &   6.599  &   6.492  &   6.537  &   6.808  &   7.042  &   7.161  &   7.029  \\
$\rm H_\beta$&   4.351  &   2.923  &   2.623  &   2.348  &   2.129  &   1.994  &   1.914  &   1.822  &   1.714  &   1.672  &   1.689  &   1.527  &   1.409  &   1.423  &   1.446  \\
Fe5015       &   4.426  &   5.159  &   5.742  &   6.026  &   5.845  &   6.006  &   5.882  &   6.279  &   6.478  &   6.329  &   6.300  &   6.486  &   6.602  &   6.817  &   6.608  \\
$\rm Mg_1$   &   0.027  &   0.052  &   0.065  &   0.076  &   0.087  &   0.095  &   0.102  &   0.107  &   0.111  &   0.114  &   0.120  &   0.124  &   0.132  &   0.129  &   0.130  \\
$\rm Mg_2$   &   0.121  &   0.165  &   0.196  &   0.217  &   0.229  &   0.242  &   0.248  &   0.264  &   0.275  &   0.274  &   0.282  &   0.295  &   0.303  &   0.306  &   0.311  \\
$\rm Mg_b$   &   2.108  &   2.741  &   3.349  &   3.669  &   3.707  &   3.837  &   3.816  &   4.149  &   4.320  &   4.239  &   4.265  &   4.554  &   4.543  &   4.776  &   4.796  \\
Fe5270       &   2.011  &   2.476  &   2.710  &   2.868  &   2.953  &   3.052  &   3.096  &   3.194  &   3.283  &   3.293  &   3.334  &   3.408  &   3.498  &   3.510  &   3.495  \\
Fe5335       &   1.802  &   2.272  &   2.506  &   2.665  &   2.758  &   2.858  &   2.904  &   3.001  &   3.081  &   3.090  &   3.134  &   3.197  &   3.278  &   3.283  &   3.271  \\
Fe5406       &   0.981  &   1.323  &   1.472  &   1.588  &   1.697  &   1.772  &   1.830  &   1.875  &   1.927  &   1.949  &   1.995  &   2.038  &   2.107  &   2.075  &   2.078  \\
Fe5709       &   0.713  &   0.873  &   0.903  &   0.938  &   1.016  &   1.037  &   1.075  &   1.053  &   1.065  &   1.095  &   1.112  &   1.121  &   1.146  &   1.101  &   1.116  \\
Fe5782       &   0.555  &   0.692  &   0.743  &   0.785  &   0.843  &   0.871  &   0.897  &   0.905  &   0.917  &   0.933  &   0.959  &   0.968  &   1.002  &   0.962  &   0.975  \\
Na D         &   2.029  &   2.537  &   2.880  &   3.087  &   3.181  &   3.331  &   3.387  &   3.566  &   3.657  &   3.675  &   3.776  &   3.854  &   4.017  &   4.050  &   4.049  \\
$\rm TiO_1$  &   0.037  &   0.037  &   0.048  &   0.052  &   0.043  &   0.046  &   0.042  &   0.053  &   0.055  &   0.049  &   0.049  &   0.051  &   0.052  &   0.057  &   0.054  \\
$\rm TiO_2$  &   0.044  &   0.048  &   0.071  &   0.080  &   0.065  &   0.071  &   0.063  &   0.085  &   0.089  &   0.077  &   0.078  &   0.082  &   0.085  &   0.093  &   0.088  \\
$\rm H\delta_A$  &   5.846  &   1.715  &   0.720  &  -0.290  &  -0.990  &  -1.488  &  -1.751  &  -2.230  &  -2.828  &  -2.802  &  -2.434  &  -3.533  &  -4.119  &  -4.180  &  -3.845  \\
$\rm H\gamma_A$  &   4.122  &  -0.863  &  -2.227  &  -3.453  &  -4.287  &  -4.871  &  -5.172  &  -5.688  &  -6.333  &  -6.245  &  -5.992  &  -7.009  &  -7.628  &  -7.652  &  -7.271  \\
$\rm H\delta_F$  &   4.168  &   2.052  &   1.541  &   1.035  &   0.705  &   0.509  &   0.405  &   0.190  &  -0.120  &  -0.103  &   0.130  &  -0.422  &  -0.638  &  -0.684  &  -0.560  \\
$\rm H\gamma_F$  &   3.912  &   1.356  &   0.585  &  -0.071  &  -0.497  &  -0.814  &  -0.979  &  -1.254  &  -1.606  &  -1.559  &  -1.457  &  -1.962  &  -2.287  &  -2.351  &  -2.126  \\

\hline
\end{tabular}
\label{lick-hs-ff}
\end{minipage}
\end{table*}

\normalsize
\begin{figure*}
\psfig{file=MF414rv2fig7.ps,height=10.5cm,width=16cm,bbllx=575pt,bblly=103pt,bburx=80pt,bbury=691pt,clip=,angle=270}
\caption{Evolution of absorption indices in the Lick/IDS system
obtained by using the fitting functions for BSPs of various
metallicity. The symbols linked by a line denote the indices
obtained by using the high resolution spectra of \citet{gon05},
and those without a line are from the low resolution spectra of
\citet{lej97,lej98}. Different symbols are for different
metallicity, from top to bottom, the metallicity $Z$ is 0.03,
0.02, 0.01 and 0.004, respectively.} \label{lick-ff-com}
\end{figure*}

\begin{figure}
\psfig{file=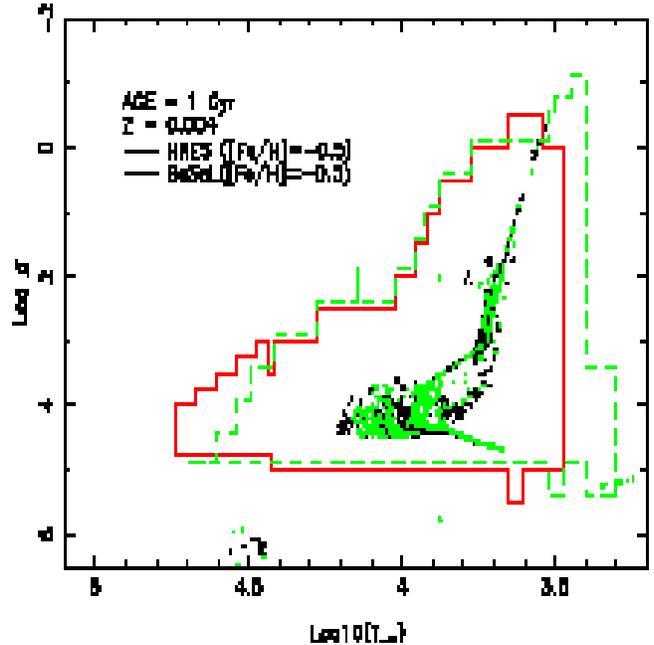,height=8.5cm,width=8.5cm,bbllx=395pt,bblly=0pt,bburx=0pt,bbury=500pt,clip=,angle=270}
\caption{Coverage of HRES (solid line) and BaSeL-2.0 (dashed line)
stellar library grids for [Fe/H]$=-0.5$. For the sake of clarity
BaSeL-2.0 library is shifted upwards by an amount of 0.1.
Over-plotted points represent the theoretical isochrone for
instantaneous burst BSPs with $Z=0.004$ and $\tau = 1$\,Gyr. Only
200,000 binary systems are plotted and those MS stars with mass $M
\la 0.7 M_{\rm \odot}$ are removed.} \label{iso-2lib1}
\end{figure}

\begin{figure}
\psfig{file=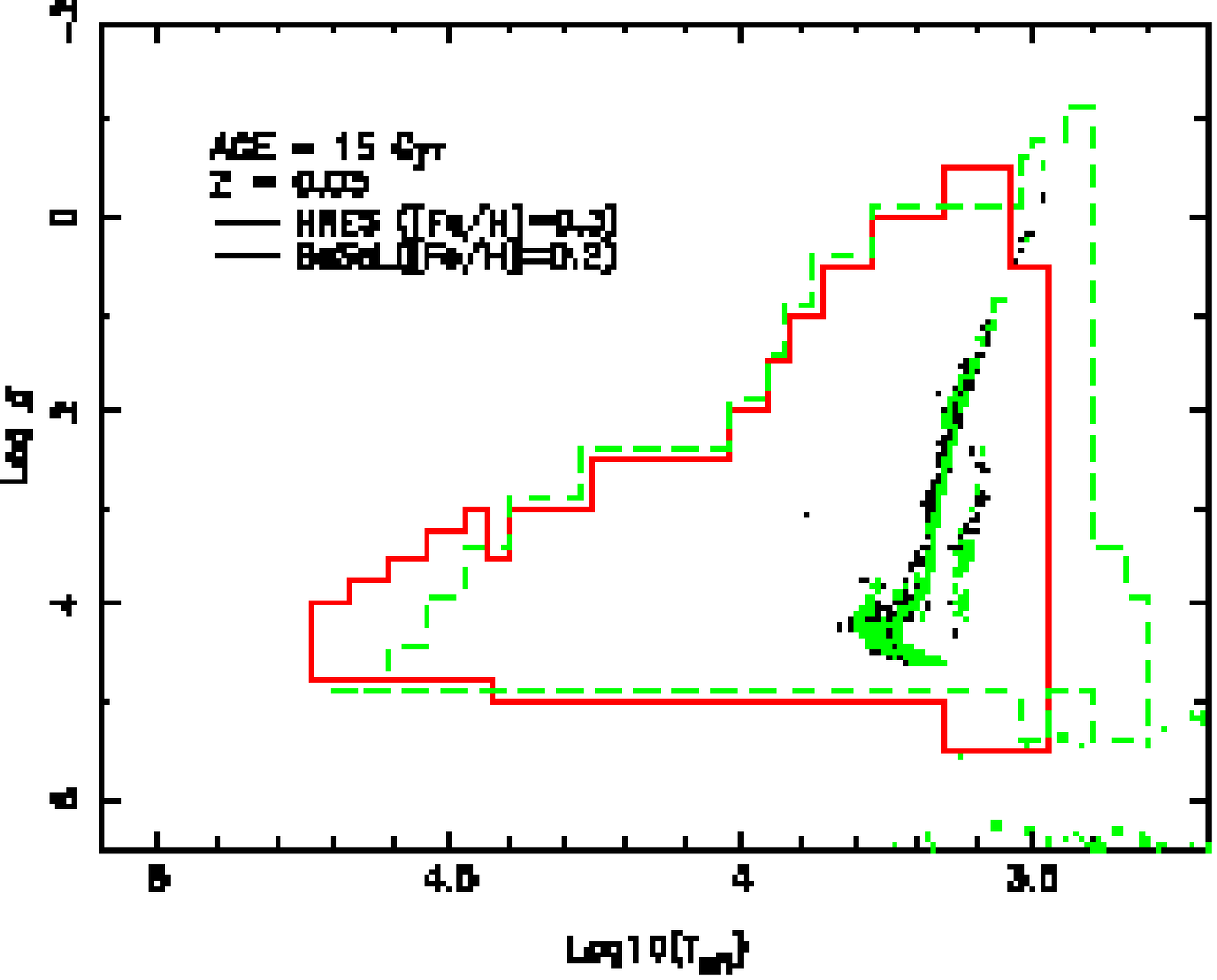,height=8.5cm,width=8.5cm,bbllx=395pt,bblly=0pt,bburx=0pt,bbury=500pt,clip=,angle=270}
\caption{Similar to Fig. \ref{iso-2lib1}, but [Fe/H]$=0.3$ for
HRES and [Fe/H]$=0.2$ for BaSeL-2.0 libraries, $Z=0.03$ and $\tau
= 15$\,Gyr for the BSPs.} \label{iso-2lib2}
\end{figure}

Using the high resolution spectra and the fitting functions of
\citet{wor94} and \citet{wor97}, we obtain 25 Lick/IDS spectral
absorption indices for the BSPs with binary interactions with four
metallicities from $Z=0.004$ to $Z=0.03$. The results are
presented in Table \ref{lick-hs-ff}. We compare them with those
obtained by Model A of Paper II, which uses low resolution stellar
spectral library of and the fitting functions, and find that the
evolution of $\rm TiO_1$ and $\rm TiO_2$ indices is not smooth for
both two studies, the discrepancy in the rest 23 Lick/IDS spectral
indices is small for metallicity $Z=0.004$ and $Z=0.02$, the
synthetic Ca4227 (index 3), Fe5015 (10), $\rm Mg_1$ (11), $\rm
Mg_b$ (13), Fe5709 (17) and Fe5782 (18) indices show much stronger
discrepancy for metallicity $Z=0.01$ and $Z=0.03$ .

In Fig. ~\ref{lick-ff-com} we show the comparison of Ca4227 (index
3), Fe5015 (10), $\rm Mg_1$ (11), $\rm Mg_b$ (13), Fe5709 (17) and
Fe5782 (18) indices between two studies. It shows that the high
spectral resolution Ca4227 (3), Fe5015 (10) and $\rm Mg_b$ (13)
indices are greater, $\rm Mg_1$ (11), Fe5709 (17) and Fe5782 (18)
indices are less than those at low resolution for BSPs with
$Z=0.03$ and $Z=0.01$, the discrepancy is greater for BSPs with
$Z=0.03$ in comparison to $Z=0.01$. This discrepancy in the
Lick/IDS spectral indices will directly influence the derived age
and metallicity of a particular population, it will lower the age
and metallicity if using high resolution Ca4227 (3), Fe5015 (10)
and $\rm Mg_b$ (13) indices, and raise the age and metallicity if
using high resolution $\rm Mg_1$ (11), Fe5709 (17) and Fe5782 (18)
indices.

To discuss whether the difference of the coverage in the
log$g-$log$T_{\rm eff}$ plane between HRES and BaSeL-2.0 libraries
would cause the discrepancies in the Lick/IDS spectral absorption
indices, in Figs. \ref{iso-2lib1} and \ref{iso-2lib2} we plot the
theoretical isochrones of the bluest (i.e., the youngest
[$\tau=1$\,Gyr] and lowest metallicity [$Z=0.004$]) and the
reddest (the oldest [$\tau=15$\,Gyr] and highest metallicity
$Z=0.03$) BSPs. In order to transform the evolutionary parameters
of stars along the isochrones to stellar flux for $Z=0.004$ BSPs
in Fig. \ref{iso-2lib1}, we need to make interpolation in the flux
grids between [Fe/H]$=-1.0$ and $-0.5$ because HRES spectral
library only contains [Fe/H] $=-1.0, -0.5, 0.0$ and 0.3. While,
the grid coverage in the log$g-$log$T_{\rm eff}$ plane between
[Fe/H] $=-1.0$ and $-0.5$ is appreciably different, and the
difference mainly concentrates in the low-temperature range.
Within this low temperature range the denser stars (such as,
low-mass MS stars) only give small contribution to the ISEDs (see
Figs. 3 and 4 of Paper II), so we only interest in the difference
of two libraries in the low-density range, in this range the grid
coverage at [Fe/H] $=-0.5$ is smaller than that at [Fe/H] $=-1.0$,
so in Fig. \ref{iso-2lib1} we plot the boundary of HRES flux grid
coverage at [Fe/H] $=-0.5$. Also we plot the boundary of BaSeL-2.0
flux grid coverage at [Fe/H] $=-0.5$, for this library the grid
coverage in the log$g-$log$T_{\rm eff}$ plane for [Fe/H] $=-1.0$
and $-0.5$ is same in the low-temperature and low-density range.
Similarly, for $Z=0.03$ BSPs in Fig. \ref{iso-2lib2}, we plot the
boundary of HRES flux grid coverage at [Fe/H] $=0.3$ and BaSeL-2.0
library at [Fe/H] $=0.2$.

Comparison of the coverage in the log$g-$log$T_{\rm eff}$ plane
for two libraries in Figs. \ref{iso-2lib1} and \ref{iso-2lib2}
shows HRES library extends to the higher temperature, in the
log$T_{\rm eff} \ge 4.43$ range covers less gravity than BaSeL-2.0
library, while the latter covers lower temperature range. Although
in the high temperature range the grid coverage of two libraries
exists discrepancies, both grid coverage are enough to the BSPs we
considered. In the lower temperature and lower gravity range, HRES
library does not cover all cool AGB stars, it would cause the
discrepancies in the Lick/IDS absorption line indices.
Additionally, the differences in the spectral shape caused by the
resolution also would cause to the discrepancies of the above
several Lick/IDS absorption line indices.

\subsubsection{Results computed from the high-resolution
spectra}
\begin{table*}
\tiny
\centering
\begin{minipage}{200mm}
\caption{Lick/IDS spectral absorption feature indices of BSPs, as
measured directly from the high-resolution ISEDs (Section 3.2.2).
}
\begin{tabular}{rrrrrrrrrrrrrrrr}
\hline
  \multicolumn{16}{c}{Age (Gyr)} \\
  &   1.00  &   2.00  &   3.00  &   4.00  &   5.00  &   6.00  &   7.00  &   8.00  &   9.00  &  10.00  &  11.00  &  12.00  &  13.00  &  14.00  &  15.00  \\
\hline
\multicolumn{16}{c}{$Z$ = 0.004} \\
$\rm CN_1$       &  -0.217  &  -0.128  &  -0.078  &  -0.052  &  -0.039  &  -0.026  &  -0.022  &  -0.017  &  -0.015  &  -0.010  &  -0.008  &  -0.011  &  -0.010  &  -0.009  &  -0.010  \\
$\rm CN_2$       &  -0.166  &  -0.081  &  -0.035  &  -0.012  &  -0.001  &   0.010  &   0.013  &   0.016  &   0.019  &   0.022  &   0.024  &   0.022  &   0.023  &   0.024  &   0.022  \\
Ca4227           &   0.328  &   0.467  &   0.574  &   0.695  &   0.763  &   0.875  &   0.917  &   0.906  &   0.985  &   1.051  &   1.081  &   1.083  &   1.145  &   1.177  &   1.172  \\
G4300            &  -0.408  &   1.347  &   2.613  &   3.630  &   4.229  &   4.837  &   5.059  &   5.061  &   5.615  &   5.858  &   5.999  &   5.770  &   5.906  &   6.188  &   6.244  \\
Fe4383           &   0.104  &   1.448  &   2.298  &   2.912  &   3.264  &   3.699  &   3.832  &   3.811  &   4.100  &   4.247  &   4.378  &   4.308  &   4.461  &   4.551  &   4.534  \\
Ca4455           &   0.229  &   0.404  &   0.542  &   0.652  &   0.712  &   0.791  &   0.809  &   0.796  &   0.843  &   0.868  &   0.883  &   0.851  &   0.895  &   0.893  &   0.890  \\
Fe4531           &   1.309  &   1.832  &   2.146  &   2.405  &   2.554  &   2.713  &   2.765  &   2.739  &   2.882  &   2.945  &   2.992  &   2.942  &   3.030  &   3.074  &   3.080  \\
Fe4668           &   0.404  &   0.741  &   0.955  &   1.112  &   1.200  &   1.307  &   1.344  &   1.305  &   1.387  &   1.394  &   1.416  &   1.344  &   1.362  &   1.367  &   1.379  \\
$\rm H_\beta$    &   6.000  &   4.490  &   3.462  &   2.917  &   2.607  &   2.323  &   2.217  &   2.081  &   2.008  &   1.916  &   1.843  &   1.885  &   1.888  &   1.765  &   1.769  \\
Fe5015           &   2.413  &   3.420  &   3.983  &   4.429  &   4.631  &   4.911  &   4.974  &   4.882  &   5.113  &   5.203  &   5.236  &   5.125  &   5.275  &   5.290  &   5.303  \\
$\rm Mg_1$       &   0.017  &   0.031  &   0.042  &   0.048  &   0.051  &   0.059  &   0.061  &   0.060  &   0.062  &   0.066  &   0.068  &   0.071  &   0.074  &   0.075  &   0.073  \\
$\rm Mg_2$       &   0.056  &   0.078  &   0.094  &   0.105  &   0.111  &   0.123  &   0.128  &   0.126  &   0.133  &   0.137  &   0.142  &   0.144  &   0.149  &   0.152  &   0.151  \\
$\rm Mg_b$       &   0.628  &   0.732  &   0.840  &   0.953  &   1.027  &   1.128  &   1.192  &   1.192  &   1.296  &   1.343  &   1.387  &   1.418  &   1.456  &   1.509  &   1.523  \\
Fe5270           &   1.023  &   1.435  &   1.666  &   1.856  &   1.951  &   2.101  &   2.134  &   2.116  &   2.234  &   2.284  &   2.323  &   2.304  &   2.383  &   2.415  &   2.418  \\
Fe5335           &   1.084  &   1.477  &   1.717  &   1.919  &   1.993  &   2.179  &   2.212  &   2.166  &   2.271  &   2.348  &   2.371  &   2.375  &   2.469  &   2.475  &   2.461  \\
Fe5406           &   0.747  &   1.052  &   1.234  &   1.365  &   1.424  &   1.557  &   1.595  &   1.555  &   1.632  &   1.680  &   1.711  &   1.724  &   1.777  &   1.790  &   1.780  \\
Fe5709           &   0.356  &   0.495  &   0.564  &   0.628  &   0.654  &   0.695  &   0.692  &   0.684  &   0.725  &   0.731  &   0.734  &   0.710  &   0.731  &   0.736  &   0.740  \\
Fe5782           &   0.257  &   0.352  &   0.415  &   0.461  &   0.472  &   0.540  &   0.547  &   0.518  &   0.527  &   0.555  &   0.559  &   0.578  &   0.602  &   0.586  &   0.564  \\
Na D             &   0.713  &   0.908  &   1.052  &   1.152  &   1.205  &   1.329  &   1.384  &   1.366  &   1.441  &   1.510  &   1.548  &   1.603  &   1.663  &   1.696  &   1.699  \\
$\rm TiO_1$      &   0.009  &   0.009  &   0.010  &   0.009  &   0.009  &   0.009  &   0.011  &   0.009  &   0.009  &   0.009  &   0.010  &   0.012  &   0.011  &   0.010  &   0.010  \\
$\rm TiO_2$      &   0.020  &   0.022  &   0.026  &   0.024  &   0.025  &   0.026  &   0.030  &   0.026  &   0.026  &   0.026  &   0.028  &   0.031  &   0.030  &   0.029  &   0.028  \\
$\rm H\delta_A$  &   9.757  &   5.890  &   3.371  &   1.853  &   0.997  &   0.131  &  -0.199  &  -0.478  &  -0.897  &  -1.266  &  -1.445  &  -1.254  &  -1.349  &  -1.657  &  -1.667  \\
$\rm H\gamma_A$  &   8.519  &   4.451  &   1.617  &  -0.240  &  -1.353  &  -2.462  &  -2.907  &  -3.168  &  -3.898  &  -4.359  &  -4.646  &  -4.384  &  -4.543  &  -5.057  &  -5.113  \\
$\rm H\delta_F$  &   6.969  &   4.834  &   3.271  &   2.334  &   1.797  &   1.283  &   1.094  &   0.874  &   0.663  &   0.441  &   0.346  &   0.491  &   0.458  &   0.253  &   0.245  \\
$\rm H\gamma_F$  &   6.460  &   4.326  &   2.712  &   1.685  &   1.066  &   0.444  &   0.196  &  -0.008  &  -0.355  &  -0.609  &  -0.771  &  -0.633  &  -0.678  &  -0.986  &  -1.010  \\

\vspace*{0.3pt} \\
\multicolumn{16}{c}{$Z$ = 0.01} \\
$\rm CN_1$       &  -0.164  &  -0.058  &  -0.027  &  -0.016  &  -0.004  &   0.008  &   0.011  &   0.017  &   0.025  &   0.024  &   0.026  &   0.026  &   0.030  &   0.034  &   0.034  \\
$\rm CN_2$       &  -0.110  &  -0.011  &   0.016  &   0.026  &   0.037  &   0.047  &   0.049  &   0.056  &   0.064  &   0.063  &   0.064  &   0.065  &   0.068  &   0.072  &   0.072  \\
Ca4227           &   0.430  &   0.710  &   0.931  &   1.037  &   1.144  &   1.272  &   1.272  &   1.403  &   1.497  &   1.546  &   1.549  &   1.591  &   1.637  &   1.654  &   1.719  \\
G4300            &   0.419  &   2.984  &   4.362  &   4.797  &   5.213  &   5.704  &   5.726  &   6.010  &   6.278  &   6.350  &   6.485  &   6.458  &   6.521  &   6.832  &   6.765  \\
Fe4383           &   0.997  &   3.022  &   3.963  &   4.364  &   4.760  &   5.182  &   5.235  &   5.592  &   5.843  &   5.930  &   5.993  &   6.071  &   6.190  &   6.336  &   6.410  \\
Ca4455           &   0.408  &   0.782  &   0.949  &   1.020  &   1.089  &   1.162  &   1.170  &   1.232  &   1.276  &   1.285  &   1.291  &   1.301  &   1.329  &   1.346  &   1.364  \\
Fe4531           &   1.953  &   2.742  &   3.086  &   3.222  &   3.363  &   3.502  &   3.510  &   3.645  &   3.737  &   3.784  &   3.805  &   3.821  &   3.883  &   3.942  &   3.979  \\
Fe4668           &   0.935  &   1.499  &   1.754  &   1.859  &   1.934  &   2.041  &   2.075  &   2.111  &   2.141  &   2.156  &   2.190  &   2.154  &   2.169  &   2.248  &   2.226  \\
$\rm H_\beta$    &   5.579  &   3.653  &   2.942  &   2.695  &   2.500  &   2.235  &   2.153  &   2.080  &   1.944  &   1.928  &   1.840  &   1.809  &   1.792  &   1.658  &   1.695  \\
Fe5015           &   3.721  &   5.172  &   5.702  &   5.903  &   6.109  &   6.296  &   6.324  &   6.506  &   6.607  &   6.671  &   6.661  &   6.646  &   6.735  &   6.787  &   6.849  \\
$\rm Mg_1$       &   0.030  &   0.056  &   0.069  &   0.075  &   0.083  &   0.090  &   0.090  &   0.099  &   0.104  &   0.107  &   0.107  &   0.109  &   0.113  &   0.113  &   0.118  \\
$\rm Mg_2$       &   0.081  &   0.122  &   0.144  &   0.155  &   0.167  &   0.179  &   0.181  &   0.193  &   0.201  &   0.208  &   0.208  &   0.211  &   0.216  &   0.219  &   0.225  \\
$\rm Mg_b$       &   0.795  &   1.115  &   1.395  &   1.540  &   1.667  &   1.795  &   1.842  &   1.951  &   2.026  &   2.140  &   2.180  &   2.193  &   2.227  &   2.319  &   2.347  \\
Fe5270           &   1.513  &   2.095  &   2.377  &   2.501  &   2.636  &   2.753  &   2.792  &   2.912  &   2.984  &   3.034  &   3.053  &   3.080  &   3.132  &   3.182  &   3.216  \\
Fe5335           &   1.493  &   2.137  &   2.457  &   2.596  &   2.748  &   2.894  &   2.917  &   3.077  &   3.168  &   3.208  &   3.184  &   3.231  &   3.288  &   3.277  &   3.355  \\
Fe5406           &   1.079  &   1.552  &   1.774  &   1.880  &   1.992  &   2.096  &   2.116  &   2.231  &   2.294  &   2.349  &   2.332  &   2.347  &   2.396  &   2.403  &   2.467  \\
Fe5709           &   0.523  &   0.694  &   0.778  &   0.804  &   0.833  &   0.858  &   0.871  &   0.890  &   0.900  &   0.899  &   0.908  &   0.912  &   0.917  &   0.939  &   0.929  \\
Fe5782           &   0.342  &   0.536  &   0.639  &   0.679  &   0.733  &   0.778  &   0.769  &   0.832  &   0.862  &   0.864  &   0.825  &   0.840  &   0.861  &   0.817  &   0.865  \\
Na D             &   0.937  &   1.336  &   1.556  &   1.673  &   1.811  &   1.915  &   1.951  &   2.086  &   2.168  &   2.262  &   2.254  &   2.284  &   2.348  &   2.380  &   2.462  \\
$\rm TiO_1$      &   0.013  &   0.017  &   0.016  &   0.016  &   0.017  &   0.017  &   0.017  &   0.018  &   0.018  &   0.021  &   0.019  &   0.018  &   0.019  &   0.019  &   0.021  \\
$\rm TiO_2$      &   0.026  &   0.036  &   0.036  &   0.038  &   0.040  &   0.041  &   0.041  &   0.043  &   0.044  &   0.048  &   0.045  &   0.043  &   0.045  &   0.044  &   0.048  \\
$\rm H\delta_A$  &   7.907  &   2.879  &   0.912  &   0.179  &  -0.603  &  -1.435  &  -1.644  &  -2.077  &  -2.639  &  -2.729  &  -2.976  &  -2.995  &  -3.181  &  -3.669  &  -3.612  \\
$\rm H\gamma_A$  &   6.638  &   1.004  &  -1.561  &  -2.476  &  -3.389  &  -4.410  &  -4.610  &  -5.149  &  -5.770  &  -5.948  &  -6.269  &  -6.274  &  -6.441  &  -7.108  &  -7.004  \\
$\rm H\delta_F$  &   6.070  &   3.089  &   1.909  &   1.503  &   1.081  &   0.613  &   0.481  &   0.310  &  -0.002  &  -0.030  &  -0.170  &  -0.160  &  -0.254  &  -0.537  &  -0.474  \\
$\rm H\gamma_F$  &   5.654  &   2.655  &   1.243  &   0.740  &   0.263  &  -0.315  &  -0.453  &  -0.693  &  -1.032  &  -1.122  &  -1.308  &  -1.316  &  -1.389  &  -1.755  &  -1.685  \\

\vspace*{0.3pt} \\
\multicolumn{16}{c}{$Z$ = 0.02} \\
$\rm CN_1$       &  -0.108  &  -0.031  &  -0.004  &   0.015  &   0.031  &   0.044  &   0.057  &   0.058  &   0.068  &   0.079  &   0.084  &   0.078  &   0.093  &   0.084  &   0.098  \\
$\rm CN_2$       &  -0.053  &   0.016  &   0.041  &   0.059  &   0.074  &   0.087  &   0.100  &   0.099  &   0.110  &   0.122  &   0.127  &   0.122  &   0.136  &   0.126  &   0.143  \\
Ca4227           &   0.579  &   1.107  &   1.425  &   1.610  &   1.774  &   1.897  &   2.061  &   2.024  &   2.155  &   2.314  &   2.355  &   2.249  &   2.412  &   2.266  &   2.540  \\
G4300            &   1.424  &   4.151  &   5.215  &   5.706  &   6.161  &   6.442  &   6.733  &   6.576  &   6.823  &   7.095  &   7.091  &   6.884  &   7.219  &   6.655  &   7.339  \\
Fe4383           &   2.107  &   4.127  &   5.158  &   5.765  &   6.278  &   6.658  &   7.107  &   7.054  &   7.420  &   7.835  &   7.966  &   7.813  &   8.252  &   7.870  &   8.550  \\
Ca4455           &   0.696  &   1.071  &   1.263  &   1.371  &   1.468  &   1.545  &   1.618  &   1.598  &   1.679  &   1.752  &   1.771  &   1.752  &   1.818  &   1.737  &   1.862  \\
Fe4531           &   2.612  &   3.370  &   3.728  &   3.913  &   4.083  &   4.209  &   4.352  &   4.316  &   4.464  &   4.608  &   4.641  &   4.611  &   4.730  &   4.574  &   4.867  \\
Fe4668           &   1.449  &   1.879  &   2.179  &   2.311  &   2.495  &   2.572  &   2.657  &   2.600  &   2.720  &   2.810  &   2.836  &   2.776  &   2.947  &   2.791  &   2.863  \\
$\rm H_\beta$    &   5.050  &   3.399  &   2.889  &   2.579  &   2.340  &   2.173  &   2.001  &   1.904  &   1.832  &   1.740  &   1.643  &   1.706  &   1.522  &   1.452  &   1.436  \\
Fe5015           &   5.113  &   6.358  &   6.933  &   7.183  &   7.461  &   7.634  &   7.811  &   7.688  &   7.943  &   8.123  &   8.147  &   8.043  &   8.240  &   7.972  &   8.298  \\
$\rm Mg_1$       &   0.048  &   0.076  &   0.094  &   0.106  &   0.116  &   0.123  &   0.133  &   0.134  &   0.142  &   0.150  &   0.155  &   0.151  &   0.161  &   0.159  &   0.170  \\
$\rm Mg_2$       &   0.110  &   0.163  &   0.195  &   0.213  &   0.230  &   0.241  &   0.258  &   0.258  &   0.270  &   0.285  &   0.291  &   0.285  &   0.298  &   0.295  &   0.312  \\
$\rm Mg_b$       &   0.995  &   1.641  &   2.041  &   2.240  &   2.465  &   2.570  &   2.770  &   2.775  &   2.874  &   3.052  &   3.136  &   3.084  &   3.182  &   3.189  &   3.366  \\
Fe5270           &   2.043  &   2.601  &   2.921  &   3.115  &   3.269  &   3.382  &   3.511  &   3.514  &   3.623  &   3.741  &   3.797  &   3.807  &   3.916  &   3.831  &   4.024  \\
Fe5335           &   2.044  &   2.809  &   3.221  &   3.469  &   3.635  &   3.752  &   3.917  &   3.901  &   3.999  &   4.140  &   4.188  &   4.126  &   4.266  &   4.159  &   4.358  \\
Fe5406           &   1.433  &   1.966  &   2.273  &   2.431  &   2.579  &   2.667  &   2.809  &   2.787  &   2.894  &   3.016  &   3.060  &   2.991  &   3.109  &   3.068  &   3.194  \\
Fe5709           &   0.705  &   0.834  &   0.896  &   0.939  &   0.966  &   0.988  &   1.000  &   1.005  &   1.019  &   1.028  &   1.038  &   1.059  &   1.072  &   1.040  &   1.083  \\
Fe5782           &   0.495  &   0.803  &   0.955  &   1.040  &   1.084  &   1.118  &   1.172  &   1.150  &   1.186  &   1.222  &   1.221  &   1.157  &   1.214  &   1.167  &   1.215  \\
Na D             &   1.250  &   1.809  &   2.142  &   2.334  &   2.499  &   2.602  &   2.772  &   2.793  &   2.909  &   3.045  &   3.115  &   3.080  &   3.209  &   3.208  &   3.364  \\
$\rm TiO_1$      &   0.013  &   0.018  &   0.022  &   0.020  &   0.024  &   0.024  &   0.027  &   0.025  &   0.029  &   0.032  &   0.032  &   0.027  &   0.030  &   0.033  &   0.031  \\
$\rm TiO_2$      &   0.029  &   0.040  &   0.048  &   0.048  &   0.053  &   0.054  &   0.060  &   0.057  &   0.063  &   0.067  &   0.067  &   0.061  &   0.066  &   0.068  &   0.068  \\
$\rm H\delta_A$  &   5.770  &   1.301  &  -0.464  &  -1.571  &  -2.637  &  -3.449  &  -4.261  &  -4.282  &  -4.956  &  -5.630  &  -5.925  &  -5.442  &  -6.476  &  -5.855  &  -6.872  \\
$\rm H\gamma_A$  &   4.411  &  -1.049  &  -3.219  &  -4.425  &  -5.576  &  -6.367  &  -7.214  &  -7.184  &  -7.830  &  -8.514  &  -8.775  &  -8.300  &  -9.282  &  -8.631  &  -9.750  \\
$\rm H\delta_F$  &   4.901  &   2.196  &   1.242  &   0.667  &   0.124  &  -0.296  &  -0.661  &  -0.703  &  -1.041  &  -1.282  &  -1.427  &  -1.136  &  -1.678  &  -1.460  &  -1.769  \\
$\rm H\gamma_F$  &   4.657  &   1.728  &   0.581  &  -0.070  &  -0.686  &  -1.090  &  -1.544  &  -1.590  &  -1.886  &  -2.209  &  -2.376  &  -2.072  &  -2.620  &  -2.433  &  -2.806  \\

\vspace*{0.3pt} \\
\multicolumn{16}{c}{$Z$ = 0.03} \\
$\rm CN_1$       &  -0.074  &  -0.010  &   0.015  &   0.031  &   0.047  &   0.061  &   0.070  &   0.080  &   0.087  &   0.088  &   0.090  &   0.103  &   0.116  &   0.113  &   0.107  \\
$\rm CN_2$       &  -0.022  &   0.035  &   0.060  &   0.076  &   0.092  &   0.106  &   0.114  &   0.125  &   0.135  &   0.136  &   0.141  &   0.152  &   0.166  &   0.163  &   0.159  \\
Ca4227           &   0.745  &   1.419  &   1.759  &   2.016  &   2.172  &   2.352  &   2.435  &   2.617  &   2.804  &   2.815  &   2.913  &   3.091  &   3.338  &   3.342  &   3.319  \\
G4300            &   2.095  &   4.810  &   5.436  &   5.918  &   6.256  &   6.427  &   6.591  &   6.653  &   6.906  &   6.780  &   6.677  &   6.974  &   7.213  &   7.204  &   6.960  \\
Fe4383           &   2.686  &   4.710  &   5.556  &   6.196  &   6.677  &   7.111  &   7.381  &   7.741  &   8.127  &   8.129  &   8.297  &   8.649  &   9.096  &   9.096  &   9.005  \\
Ca4455           &   0.851  &   1.219  &   1.358  &   1.476  &   1.570  &   1.637  &   1.684  &   1.742  &   1.823  &   1.832  &   1.833  &   1.918  &   1.977  &   1.980  &   1.989  \\
Fe4531           &   2.923  &   3.677  &   3.947  &   4.197  &   4.379  &   4.508  &   4.602  &   4.730  &   4.896  &   4.901  &   4.935  &   5.080  &   5.213  &   5.225  &   5.245  \\
Fe4668           &   1.618  &   2.022  &   2.184  &   2.314  &   2.400  &   2.468  &   2.515  &   2.525  &   2.622  &   2.522  &   2.458  &   2.514  &   2.523  &   2.539  &   2.503  \\
$\rm H_\beta$    &   4.414  &   3.086  &   2.663  &   2.419  &   2.244  &   2.047  &   1.908  &   1.807  &   1.782  &   1.783  &   1.761  &   1.628  &   1.503  &   1.441  &   1.548  \\
Fe5015           &   5.662  &   6.810  &   7.177  &   7.510  &   7.758  &   7.914  &   8.010  &   8.149  &   8.388  &   8.361  &   8.363  &   8.549  &   8.675  &   8.635  &   8.745  \\
$\rm Mg_1$       &   0.058  &   0.087  &   0.107  &   0.122  &   0.133  &   0.145  &   0.151  &   0.162  &   0.170  &   0.171  &   0.179  &   0.185  &   0.195  &   0.196  &   0.197  \\
$\rm Mg_2$       &   0.126  &   0.184  &   0.217  &   0.242  &   0.260  &   0.279  &   0.291  &   0.306  &   0.321  &   0.322  &   0.332  &   0.341  &   0.357  &   0.361  &   0.362  \\
$\rm Mg_b$       &   1.114  &   1.865  &   2.270  &   2.581  &   2.774  &   3.001  &   3.163  &   3.294  &   3.485  &   3.476  &   3.577  &   3.655  &   3.806  &   3.965  &   3.934  \\
Fe5270           &   2.254  &   2.812  &   3.057  &   3.257  &   3.444  &   3.578  &   3.684  &   3.764  &   3.874  &   3.926  &   3.994  &   4.084  &   4.204  &   4.212  &   4.236  \\
Fe5335           &   2.372  &   3.221  &   3.606  &   3.858  &   4.061  &   4.245  &   4.339  &   4.465  &   4.563  &   4.621  &   4.734  &   4.814  &   5.011  &   4.923  &   4.927  \\
Fe5406           &   1.598  &   2.137  &   2.418  &   2.634  &   2.773  &   2.929  &   3.011  &   3.134  &   3.249  &   3.239  &   3.317  &   3.379  &   3.505  &   3.496  &   3.518  \\
Fe5709           &   0.764  &   0.872  &   0.882  &   0.903  &   0.949  &   0.956  &   0.983  &   0.962  &   0.972  &   0.989  &   0.983  &   1.000  &   1.003  &   1.010  &   1.010  \\
Fe5782           &   0.642  &   1.023  &   1.184  &   1.271  &   1.329  &   1.393  &   1.405  &   1.463  &   1.481  &   1.486  &   1.531  &   1.545  &   1.621  &   1.530  &   1.537  \\
Na D             &   1.463  &   2.069  &   2.418  &   2.686  &   2.889  &   3.083  &   3.212  &   3.362  &   3.489  &   3.520  &   3.646  &   3.720  &   3.878  &   3.892  &   3.935  \\
$\rm TiO_1$      &   0.011  &   0.014  &   0.019  &   0.024  &   0.023  &   0.027  &   0.028  &   0.032  &   0.036  &   0.032  &   0.033  &   0.034  &   0.034  &   0.037  &   0.039  \\
$\rm TiO_2$      &   0.028  &   0.036  &   0.045  &   0.053  &   0.055  &   0.060  &   0.062  &   0.069  &   0.075  &   0.070  &   0.072  &   0.073  &   0.075  &   0.076  &   0.079  \\
$\rm H\delta_A$  &   4.247  &   0.046  &  -1.575  &  -2.688  &  -3.620  &  -4.469  &  -5.043  &  -5.666  &  -6.245  &  -6.228  &  -6.352  &  -7.280  &  -8.190  &  -8.224  &  -7.829  \\
$\rm H\gamma_A$  &   2.732  &  -2.461  &  -4.135  &  -5.357  &  -6.288  &  -7.074  &  -7.674  &  -8.161  &  -8.744  &  -8.566  &  -8.621  &  -9.398  & -10.160  & -10.343  &  -9.848  \\
$\rm H\delta_F$  &   3.894  &   1.399  &   0.540  &  -0.007  &  -0.456  &  -0.857  &  -1.126  &  -1.399  &  -1.590  &  -1.567  &  -1.530  &  -2.042  &  -2.412  &  -2.401  &  -2.218  \\
$\rm H\gamma_F$  &   3.738  &   1.017  &   0.080  &  -0.548  &  -0.997  &  -1.452  &  -1.768  &  -2.032  &  -2.264  &  -2.135  &  -2.158  &  -2.536  &  -2.924  &  -3.030  &  -2.732  \\

\hline
\end{tabular}
\label{lick-hs-th}
\end{minipage}
\end{table*}

\begin{figure*}
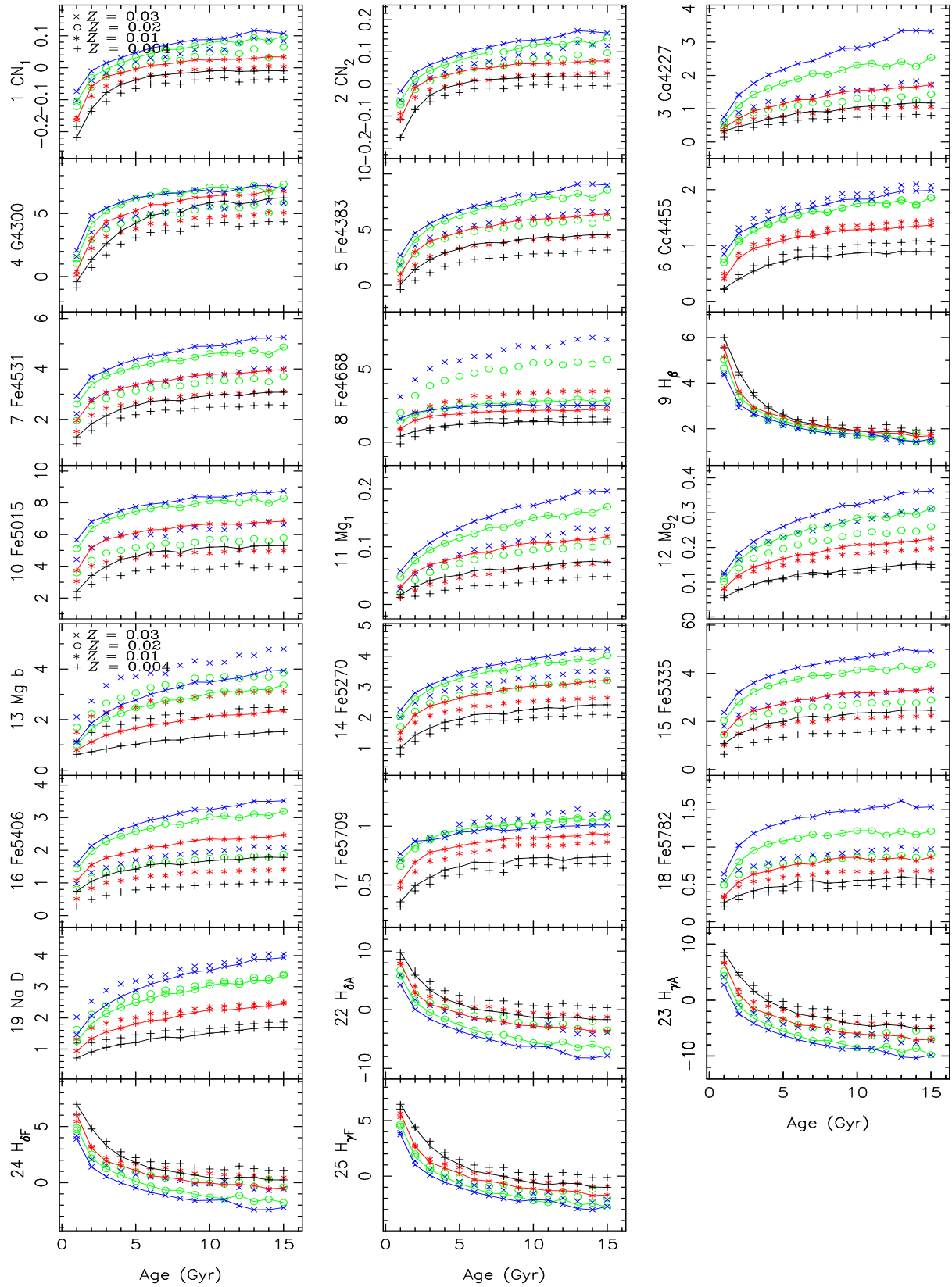

\psfig{file=MF414rv2fig101.ps,height=10.50cm,width=16cm,bbllx=533pt,bblly=103pt,bburx=80pt,bbury=691pt,clip=,angle=270}
\psfig{file=MF414rv2fig102.ps,height=11.35cm,width=16cm,bbllx=575pt,bblly=103pt,bburx=80pt,bbury=691pt,clip=,angle=270}
\caption{Absorption indices in the Lick/IDS system computed
directly from the high-resolution ISEDs (symbols + line) and by
using the fitting functions (only symbols) for BSPs of various
metallicity. The symbols have the same meaning as in Fig.
\ref{lick-ff-com}.} \label{lick-com-th-ff}
\end{figure*}

To compare theoretical results with observations, the Lick/IDS
absorption indices also are measured directly from the
high-resolution ISEDs using the passband definitions of
\citet{wor94} and \citet{wor97}, and are presented in Table
\ref{lick-hs-th}. In Fig. \ref{lick-com-th-ff} we compare them
(except for TiO$_1$ and TiO$_2$) with those obtained by using the
fitting functions of \citet{wor94} and \citet{wor97} (generated
from Table \ref{lick-hs-ff}), and find that Ca4455 (index 6,
almost equal for $Z=0.02$), Fe4668 (index 8, almost equal for
$Z=0.004$), Mg$\rm _b$ (13) and Na D (19) are bluer than the
corresponding ones obtained by using the fitting function for all
metallicities. For Fe5709 (17) the discrepancy becomes from
negative ($Z=0.03$) to positive ($Z=$0.01 and 0.004). Other
indices are redder than the values from Table \ref{lick-hs-ff} for
all metallicities.

Comparing the discrepancies in the Lick/IDS spectra indices caused
by the difference in the metallicity, we find that the
discrepancies in Ca4455 (index 6), Mg$_2$ (12), Mg$_b$ (13) and Na
D (19) indices introduced by using the different computation
method are smaller.

\section{SUMMARY AND CONCLUSIONS}
We have simulated realistic stellar populations composed of 100
per cent binaries by producing $1 \times 10^6$ binary systems
using a Monte Carlo technique. Using the EPS method we present the
high spectral resolution ($0.3\, \rm \AA$) ISEDs over the
wavelength range $3000 - 7000\, \rm \AA$ and 25 Lick/IDS spectral
absorption indices, for an extensive set of instantaneous burst
BSPs with binary interactions over a large range of age and
metallicity: $1 \leq \tau \leq 15$\,Gyr and $0.004 \leq Z \leq
0.03$. This set of high spectral resolution ISEDs fully satisfies
the need of studying the formation and evolution of galaxy by
analyzing the data of modern spectroscopic galaxy surveys, and can
be available on request.

By comparing the synthetic continuum at high and low resolution,
we show that there is a good agreement for BSPs with $Z=0.02$ and
smaller discrepancy for BSPs with non-solar metallicity. And, the
comparison of the Lick/IDS spectral indices at low and high
resolution, both of which are obtained by using the fitting
functions, shows that the high resolution Ca4227, Fe5015 and
Mg$_{\rm b}$ indices are redder, Mg$_1$, Fe5709 and Fe5782 indices
are bluer than those at low resolution for BSPs with $Z=0.01$ and
$0.03$. The high spectral Ca4227, Fe5015 and Mg$_{\rm b}$ indices
act to lower the age and metallicity, and the high resolution
Mg$_1$, Fe5709 and Fe5782 indices will raise the age and
metallicity.

At high resolution we compare the Lick/IDS spectral absorption
indices obtained by using the fitting functions with those
measured directly from the synthetic spectra, and find that
Ca4455, Fe4668, Mg$_b$ and Na D indices are redder for all
metallicities, Fe5709 is redder at $Z=0.03$ and becomes to be
bluer at $Z=$0.01 and 0.004, other indices are bluer for all
metallicities than the corresponding values computed from the
high-resolution ISEDs.

\section*{acknowledgements}
We acknowledge the generous support provided by the Chinese
Natural Science Foundation (Grant Nos 10303006, 10273020 \&
10433030), by the Chinese Academy of Sciences (KJCX2-SW-T06) and
by the 973 scheme (NKBRSF G1999075406). We are deeply indebted to
Dr. Lejeune for making his BaSeL-2.0 model available to us. We
also thanks to the referee for suggestions that have improved the
quality of this manuscript.

{}

\bsp
\label{lastpage}
\end{document}